\documentclass[10pt,journal,compsoc]{IEEEtran}

\usepackage{url}
\usepackage{amsmath}
\usepackage{xspace}
\usepackage{amssymb}
\usepackage{graphicx}
\usepackage{caption}
\usepackage{multirow}
\usepackage{subfigure}
\usepackage[ruled,vlined, linesnumbered]{algorithm2e}
\usepackage{epstopdf}
\usepackage{enumitem}
\usepackage{tgtermes}
\usepackage{booktabs}
\usepackage{array}
\usepackage[numbers]{natbib}
\usepackage{soul, color}
\definecolor{uscgold}{rgb}{1.0, 0.8, 0.0}
\definecolor{selectiveyellow}{rgb}{1.0, 0.73, 0.0}

\newcommand{\tabincell}[2]{\begin{tabular}{@{}#1@{}}#2\end{tabular}}
\newcommand{\hlgol}[1]{{\sethlcolor{selectiveyellow}\hl{#1}}}
\newcommand{\hlred}[1]{{\sethlcolor{red}\hl{#1}}}

\newcommand{\eat}[1]{}
\newcommand{\paratitle}[1]{\vspace{1.5ex}\noindent \textbf{#1}}
\newcommand{\ie}{\emph{i.e.,}\xspace}
\newcommand{\eg}{\emph{e.g.,}\xspace}

\newcommand{\FCAN}{\textsc{CARL}\xspace}

\begin{document}
\title{A Context-Aware User-Item Representation Learning for Item Recommendation}
\author{Libing Wu, Cong Quan, Chenliang Li$^*$\thanks{$^*$Chenliang Li is the corresponding author.}, Qian Wang, Bolong Zheng
\IEEEcompsocitemizethanks{
\IEEEcompsocthanksitem L. Wu, Q. Cong, C. Li, Q. Wang, are with State Key Lab of Software Engineering, School of Computer Science, Wuhan University, China. E-Mail: \{wu,quancong,cllee,qianwang\}@whu.edu.cn
\IEEEcompsocthanksitem B. Zheng is with School of Data and Computer Science, Sun Yat-Sen University, China. E-Mail: zblchris@gmail.com}
}

\markboth{IEEE Transactions on Knowledge and Data Engineering, Submission 2017}%
{}

\IEEEcompsoctitleabstractindextext{%
\begin{abstract}
Both reviews and user-item interactions (\ie rating scores) have been widely adopted for user rating prediction. However, these existing techniques mainly extract the latent representations for users and items in an independent and static manner. That is, a single static feature vector is derived to encode her preference without considering the particular characteristics of each candidate item. We argue that this static encoding scheme is difficult to fully capture the users' preference. In this paper, we propose a novel \textbf{c}ontext-\textbf{a}ware user-item \textbf{r}epresentation \textbf{l}earning model for rating prediction, named \FCAN. Namely, \FCAN derives a joint representation for a given user-item pair based on their individual latent features and latent feature interactions. Then, \FCAN adopts Factorization Machines to further model higher-order feature interactions on the basis of the user-item pair for rating prediction. Specifically, two separate learning components are devised in \FCAN to exploit review data and interaction data respectively: \textit{review-based feature learning} and \textit{interaction-based feature learning}. In review-based learning component, with convolution operations and attention mechanism, the relevant features for a user-item pair are extracted by jointly considering their corresponding reviews. However, these features are only reivew-driven and may not be comprehensive. Hence, interaction-based learning component further extracts complementary features from interaction data alone, also on the basis of user-item pairs. The final rating score is then derived with a dynamic linear fusion mechanism. Experiments on five real-world datasets show that \FCAN achieves significantly better rating predication accuracy than existing state-of-the-art alternatives. Also, with attention mechanism, we show that the relevant information in reviews can be highlighted to interpret the rating prediction.
\end{abstract}

}

\maketitle

\IEEEdisplaynotcompsoctitleabstractindextext
\IEEEpeerreviewmaketitle

\section{Introduction}\label{sec:intro}

Many social media websites and ecommerce systems allow users to write textual reviews to express their personal opinions towards the purchased items, along with a rating score indicating their preferences. The rich information covered in the textual reviews can reveal the characteristics of the items and also the preference of each individual user. Exploiting textual reviews has been proven to be effective for better rating prediction performance. Moreover, it is beneficial in alleviating the data sparsity and cold-start issues for recommender systems. Many existing works utilize both the review data and user interaction data to further enhance recommendation performance~\cite{kdd11:ctr,recsys13:hfht,aaai14:bao,ijcai16:rblt}.

Prior works resort to latent Dirichlet allocation (LDA)~\cite{jmlr03:lda} or non-negative matrix factorization (NMF)~\cite{nips00:nmf} to derive latent features over the reviews~\cite{kdd11:ctr,recsys13:hfht,aaai14:bao,ijcai16:rblt}. These techniques achieve better recommendation performance than conventional latent models solely based on the user-item interactions (\ie rating scores). However, one intrinsic limitation within these techniques is the bag-of-words (BOW) representation for review processing. That is, the semantic contextual information encoded in local word context is ignored.

\begin{figure*}
\centering
\includegraphics[width=.90\linewidth]{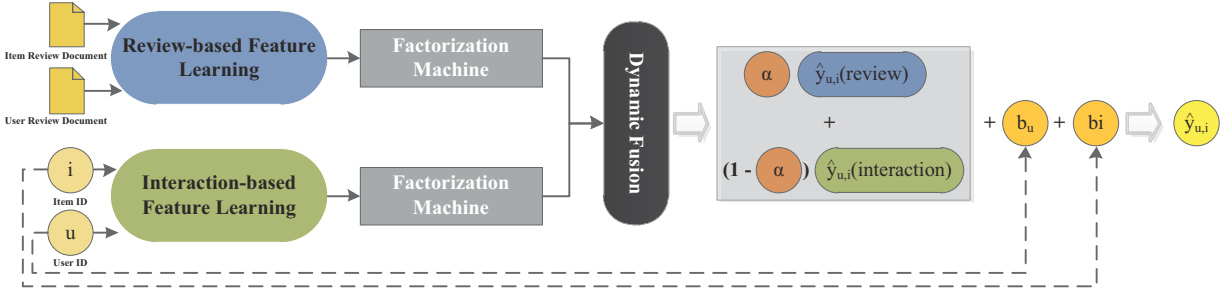}
\caption{The architecture of \FCAN.}\label{fig:fcan}
\end{figure*}

Recently, neural networks based models have been proposed to derive the latent features from the reviews for rating prediction~\cite{recsys16:convMf,wsdm17:deepconn,recsys17:transnet}. In these works, the convolutional neural network (CNN) architecture is employed to extract the user or item latent features from the corresponding user and item reviews\footnote{A user review refers to all reviews written by the user. Similarly, an item review refers to all reviews written for it.} respectively. By using dense word embedding representations and a local window to capture the contextual information, the CNN-based techniques facilitate a better semantic understanding of reviews, leading to significant improvement over the existing BOW-based methods for rating prediction. Nevertheless, people find that it is difficult to understand the features extracted by neural networks and therefore limit the interpretability of recommender system. In order to address this problem,  Seo and Huang~\cite{recsys17:dattn} incorporate attention mechanism to render a more interpretable model.

However, either the existing BOW-based methods, CNN-based methods or Attention-based methods derive user and item latent feature vectors from the reviews in an independent and static manner. A single static feature vector is assigned for a user to estimate her rating score on a candidate item whose feature vector is also static without considering the preference of the user. It is intuitive that not all words in the reviews written by a user is relevant to her rating on a particular item. For example, the critical defects mentioned by a user in her comment for an inferior-quality product would be useless to guess her preference on many other products. Identifying the relevant semantic information by jointly considering both the reviews of the user and item could be a new avenue to improve the prediction accuracy. Inspired by this idea, we further take a close look at conventional collaborative filtering (CF) techniques, \ie mainly factor learning techniques (\eg matrix factorization). It is surprising that similar observations are made. Existing CF methods mainly derive static latent feature vectors for each user and item by matching their past interaction data. Then the rating prediction is calculated by considering only the linear combination of the latent featuers through a dot-product (DP) operation. It is expected that learning a representation for a user-item pair based on the individual characteristics and their interactions together would beneift the rating predication. Here, we term this feature learning for a user-item pair as \textit{context-aware} feature learning. We argue that context-aware information extracted by modeling the latent feature interactions between users and items is more expressive and discriminative to understand users' rating behaviors. Moreover, it is easy to further facilitate the modeling of the higher-order feature interactions over the context-aware latent features for a user-item pair, and thus leads to better prediction performance.

To this end, in this paper, we propose a \textbf{c}ontext-\textbf{a}ware user-item \textbf{r}epresentation \textbf{l}earning model for rating prediction, integrating information from both textual reviews and rating scores, named \FCAN. Figure~\ref{fig:fcan} illustrates the overall network architecture for \FCAN. \FCAN consists of two independent feature learning components: \textit{review-based feature learning} and \textit{interaction-based feature learning}. In review-based feature learning, \FCAN utilizes the convolution operations and an attention mechanism to highlight the relevant semantic information by jointly considering the reviews written by a user, and the reviews written for an item. Then, an abstracting layer is employed to derive the latent feature vectors for the user and the item respectively. The representation of the user-item pair is then constructed by further considering the latent feature interactions. Note that the semantic information contained in reviews could only serve as a partial reflection of the user rating behavior. Existing interaction-based latent models have delivered promising rating prediction performance. We believe that the interaction data alone could provide complementary knowledge to complete the picture. Therefore, in interaction-based feature learning, \FCAN constructs another representation for a user-item pair based on another set of feature vectors and their interactions. \FCAN feeds the latent representation learnt by each component into a Factorization Machine (FM)~\cite{icdm10:fm} to further model higher-order feature interactions for rating prediction. Then, a novel dynamic weighted linear fusion is devised to perform the final rating prediction.

We conduct extensive experiments over five real-world datasets with textual reviews. The experimental results validate that the two feature learning components complement each other and result in better prediction performance when being fused. Also, the proposed \FCAN significantly outperforms the existing state-of-the-art alternatives across five datasets.
In summary, the contributions made in this paper are listed as follows:

\begin{itemize}
\item[$\bullet$] We propose a neural networks based context-aware user-item representation learning model for rating prediction. The model derives latent represenations on the basis of user-item pairs instead of learning a static user/item latent representation for rating prediction. To the best of our knowledge, \FCAN is the first work that learns representations for each user-item pair based on their individual characteristics and their interactions together by exploiting both textual reviews and user-item interaction data.

\item[$\bullet$] With neural attention mechanism introduced in review-based feature learning, we can identify the relevant information based on semantic interaction between a user review document and an item review document. In this sense, we are the first to extract context-aware information from reviews with neural networks techniques. The attention mechanism can also facilitate the explainability of the recommendation. Moreover, a simple but novel dynamic linear fusion strategy is proposed in \FCAN to aggregate the evidences from the two components for final prediction.

\item[$\bullet$] Through extensive experiments conducted on five real-world datasets, the results demonstrate that our proposed \FCAN consistently achieves better rating prediction accuracy than existing state-of-the-art alternatives, including slap-up BOW-based methods, CNN-based methods and Attention-based methods. Further case studies demonstrate that the words highlighted by \FCAN in reviews are very meaningful and uncover users' specific preference towards an item of interest, which helps to improve the explainability for the recommendation. 
\end{itemize}

The remainder of this paper is organized as follow: related works are reviewed in Section~\ref{sec:related}. Secion~\ref{sec:alg} introduces the overall framework of \FCAN in detail. The experimental settings and results are presented in Section~\ref{sec:exp}. Finally Section~\ref{sec:con} concludes the paper and discusses the future works.

\section{Related Work}\label{sec:related}
Our work is related to the studies of interaction based collaborative filtering, text based rating prediction, deep neural networks based recommendation and context based feature learning. Therefore, in this section, we briefly review the relevant literatures in these areas.

\subsection{Interaction based Collaborative Filtering}
The interaction based recommendation methods are mainly based on Collaborative Filtering (CF) techniques~\cite{advai09:survyCF}, which aim to represent users and items with static latent feature vectors. Matrix Factorization (MF) is the most popular technique in this line of literatures. Basic MF models, such as \cite{nips07:pmf,computer09:mf} try to learn users' and items' latent features purely by matching the user-item interaction (\ie binary indicators or user-item rating scores) matrix with dot-product (DP) operation. The rating prediction is then calculated also by using DP operation with the derived user/item latent features for a given user-item pair. Plenty of works try to enhance the performance of MF by modeling more information based on the user-item interactions. For example, \cite{computer09:mf} introduces user and item biases into MF. \cite{kdd08:neighbor1} integrates neighborhood modeling into MF. It assumes that a user's rating on an item is formed not only by the latent characteristics of the user-item pair, but also the user's rating behaviors on the other items. All these methods have been validated to outperform the vanilla MF model in many domains. However, all these MF based models use DP operation as their rating predictor. One inherent limitation of DP operation is that latent features are independent of each other. That is, DP only enables linear combination of latent features without considering higher-order feature interaction. It is validated that the performance of existing MF based mothods is hindered by this strong constraint~\cite{www17:ncf}.

With the tremendous successes of neural networks in many fields, some researchers turn to using neural networks to learn users' rating behaviors. \cite{icml07:rbmcf} proposes a restricted Boltzmann machines (RBM) based model for rating prediction. It applies separate share-parameters RBMs, each of which has visible softmax units for the rated item, to model the interaction history of each user. Hence, when two users have similar rating behavior, the prediction for them is similar as well. Neighborhood information of users and items is then utilized to extend the RBM model for better prediction performance~\cite{icml13:cfrbm}. The principle underlying these two pioneer works is still conventional user-based and item-based collaborative filtering. \cite{www17:ncf} presents a general deep neural framework for collaborative filtering with implict feedback. The proposed NeuMF takes the static user and item feature vectors as input, and calculate the rating score by replacing the DP operation with a neural atchitecture. The empirical study shows that NeuMF achieves superior performance than the existing latent factor learning techniques. NeuMF can model latent feature interactions between users and items through a deep MLP architecture and non-linearities. However, higher-order feature interactions still can not be well captured by NeuMF. In contrast, the proposed \FCAN derives a context-aware representation for a user-item pair and uses FM to model the higher-order latent feature interactions for rating prediction. Both complex interactions between users and items as well as the higher-order latent feature interactions are well captured by \FCAN to model the users' rating behaviors. Moreover, since NeuMF is devised for collaborative filtering with implicit feedback, its applicability to rating predication as well as its incorporation with review data is still unknown.

\subsection{Rating Prediction from Text}
Although CF methods~\cite{computer09:mf,icml07:rbmcf,is04:itembasedrec} based on user-item interactions are prevalent in the past decades, they have two obvious limitations. Firstly, the prediction accuracy of most CF methods drop significantly when the data is sparse. Secondly, they are incapable of handling new users and items (\ie cold-start issue). Textual information, \eg users' reviews, item description or labels, is the most popular auxiliary information available in many recommender systems. Consequently, exploiting textual information to address these inherent limitations has become a hot research topic.

Some researchers~\cite{recsys13:hfht,kdd11:ctr,ijcai16:rblt,recsys14:rmr,aaai14:bao} propose to employ topic modeling techniques to learn latent topic factors from text. HFT~\cite{recsys13:hfht} and CTR~\cite{kdd11:ctr} employ a LDA-like technique to exploit latent topics from review text. RBLT~\cite{ijcai16:rblt} employs similar techniques to uncover topic features from rating-boost review text as latent factors. The authors assume that more recommendable features would be contained in a higher-rating review. Thus, they construct the rating-boost review text of an original review by repeating the review $r$ times, where $r$ is the rating score associated with it. In this sense, topic models like LDA can easily extract these preferred features as latent topic factors. These latent topic fators are later integrated into a matrix factorization (MF) framework to derive item preferences. RMR~\cite{recsys14:rmr} uses similar technique to derive topic factors from text but they use Gaussian mixtures to model ratings instead of MF framework. TopicMF~\cite{aaai14:bao} uses MF technique to jointly model the interaction data and topics from review text. In their transform function, they use linear combination of the latent factors of users and items to transform the latent topic in the reviews. CDL \cite{kdd15:cdl} tightly couples SADE \cite{jmlr10:sdae} over the text information and PMF \cite{nips07:pmf} for the implicit rating matrix. The deep neural structure enables CDL to learn interpretable latent factors from the text. These methods outperform the models which solely rely on user-item interaction data. However, these aforementioned methods all belong to the category of bag-of-words (BOW) models which ignore the word order and the local context information. Hence, much concrete information in the form of phrases and sentences is lost through this coarse-grained text processing strategy.

To tackle this limitation, several methods endowing the MF framework with the neural treatment~\cite{ijcai16:cmle,recsys16:convMf} are proposed. CMLE~\cite{ijcai16:cmle} leverages an embedding based model to integrate word embedding model with standard MF model to accommodate the contextual information for the words in the reviews. An CNN based neural network model (named ConvMF) is proposed by \cite{recsys16:convMf}. ConvMF utilizes CNN network to obtain better latent semantic representations from textual reviews by considering the word order and the local context. ~\cite{wsdm17:deepconn} uses a parallel CNN models (named DeepCoNN) to separately derive the latent features of users and items based on their reviews. Then they concatenate the latent features of the corresponding user and item and feed the resultant vector into a Factorization Machine for rating prediction. TransNets~\cite{recsys17:transnet} extends DeepCoNN by adding an additional layer (target network) to learn the representation of target user-target item review at training time and used the learnt represention to regularize the output of source network. The source network therefore mimics latent representation of target review, yet is not available at test time. TransNets gains improvement in rating prediction against DeepCoNN. These recent methods perform better than the existing BOW based methods, and are proven to be effective in alleviating cold-start and data-sparsity issues. Despite these significant improvements on recommendation performances, these works solely derive latent feature vectors of users and items in a static and separate manner, which neglect the diverse and complex interactions between users and items.

\subsection{Attention-based Deep Recommender System}
As discussed above, neural networks based techniques have been widely applied to recommender systems. However, it is difficult to output meaningful patterns to help interpret the recommendation decisions due to the well-known black-box property. Recently, several works have been proposed to discriminate the importance of each latent features or factors to enhance recommendation accuracy, drawing on the attention mechanism recently proposed in neural networks area~\cite{corr14:attention}. With attention mchanism, we can identify the important words from textual auxiliary information and therefore provide a way to offer semantic interpretation for recommendations. D-attn~\cite{recsys17:dattn} combines local and global attentions on review text and produces weighted text. The weighted text is later passed to CNN model to derive better learnt representation of users and items. In multimedia recommendation, in order to better characterize users' preference, ACF~\cite{sigir17:acf} employs two attentive modules which learn to select informative components of multiple items and representative items from users' purchase records, respectively. ACF incorporates the attentive modules into classic CF models with implicit feedback~\cite{icdm08:impltrec,sigir16:fmif}. Other than applying attention to derive the importance of item components, DAMD~\cite{kdd17:dadm} leverages attention model to adaptively incorporate multiple prediction models based on their suitabilities for article recommendation. Though these existing attention-based recommender systems improve recommendation performances and also interpretability of recommender system , they also neglect the diverse and complex interactions between users and items. Here, we also utilize attention mechanism to identify the relevant semantic information by jointly consider the reviews of the user and item together. Differing from all the text based methods and attention based methods mentioned above, the proposed \FCAN extracts the latent features for a user-item pair based on their individual characteristics and their interactions together. The incorporation of FM further models the higher-order latent feature interactions for better rating prediction. Our experimental results show that \FCAN achieves much better rating prediction performance than existing state-of-the-art alterantives. 

\subsection{Context-based Features Learning}

Attention-based neural networks have been shown to be effective in improving performance in many tasks~\cite{corr14:attention,tacl16:yin,acl16:wang,aaai16:wan}. Many recent works have utilized the attention mechanism to learn context-aware latent representations from textual information. For example, several works propose to utilize an attention layer to learn context-based representation for question and answer matching~\cite{tacl16:yin,acl16:wang,aaai16:wan,corr16:attention}. CANE~\cite{acl17:cane} uses similar structure to learn dynamic network embeddings and yields better performance than the static embedding based methods. Similar to these techniques, \FCAN is devised to jointly infer the context-aware latent features based on both the review data and user-item rating scores. To the best of our knowledge, we are the first to introduce context-aware representation learning on the basis of user-item pairs for rating prediction.

\section{The Proposed Model}\label{sec:alg}

In this section, we present \FCAN, a \textbf{c}ontext-\textbf{a}ware user-item \textbf{r}epresentation \textbf{l}earning model for item recommendation. As demonstrated in Figure~\ref{fig:fcan}, \FCAN is devised to estimate the personalized rating score for a new user-item pair by exploiting two heterogenous information sources:  \textit{item reviews} written by users and \textit{user-item interaction matrix}. Hence, \FCAN consists of two independent feature learning components: \textit{review-based feature learning} and \textit{interaction-based feature learning} . In the following, we first detail the rating prediction framework for \FCAN, followed by the description about the two learning components.

\subsection{Rating Prediction Framework}\label{ssec:overall}
The dot-product (DP) operation is often used by the existing works for rating prediction~\cite{nips07:pmf,wsdm17:deepconn,ijcai16:cmle}. However, DP holds a strong constraint that the latent dimensions are independent of each other. That is, each dimension in a latent user vector could only interact with the corresponding dimension in the latent item vector. This independence constraint is incapable of learning complex user-item rating behaviors through the higher-order feature interactions. Since the proposed \FCAN derives a context-aware representation for each user-item pair, it is desired to model higher-order latent feature interactions for better understanding the rating behaviors. Hence, in this work, we choose Factorization Machine (FM)~\cite{icdm10:fm} to calculate the rating score. Specifically, given a latent feature vector learnt for a user-item pair, denoted as $\mathbf{z}_{u,i}$, FM calculates the corresponding rating score as follows:
\begin{align}
\hat{y}_{u,i}(\mathbf{z}_{u,i}) &= m_0 + \mathbf{m}^\mathrm{T}\mathbf{z}_{u,i} + \frac{1}{2}\mathbf{z}_{u,i}^\mathrm{T}\mathbf{M}\mathbf{z}_{u,i} \label{eqn:fm}\\
\mathbf{M}_{j,k}&=\mathbf{v}_j^\mathrm{T}\mathbf{v}_k,j\neq k\label{eqn:fmM}
\end{align}
where $m_0$ is the global bias, $\mathbf{m}$ is the coefficent vector for latent feature vector $\mathbf{z}_{u,i}$, $\mathbf{M}$ is the weight matrix for second-order interactions and its diagonal elements are $0$ (\ie $\mathbf{M}_{j,j}=0$), $\mathbf{v}_j,\mathbf{v}_k\in \mathbb{R}^v$ are the $v$-dimensional latent vectors associated with dimension $j$ and $k$ of $\mathbf{z}_{u,i}$. From Equation~\ref{eqn:fm}, we can see that both first (\ie $\mathbf{m}$) and second-order (\ie $\mathbf{M}$) feature interactions are utilized for rating prediction. FM is also applied for rating prediction in other studies like~\cite{wsdm17:deepconn,kdd08:neighbor1}. We believe capturing higher-order feature interactions is important for modeling complex user-item rating behaviors. We have also tried using other rating prediction formulas like linear regression (LR) and multiple-layer perception (MLP)~\cite{www17:ncf}. However, our results show that FM yields better prediction accuracy than LR and MLP (ref. Section~\ref{ssec:FCANAnalysis}).

In reality, rating behaviors contain multiple inherent tendencies, known as bias. For example, some users tend to rate a higher score for all items. And some items are likely to recieve higher ratings from all users~\cite{computer09:mf}.~\cite{lecture12:recommend} has proven that incorporating user and item biases can accommodate the rating variations well and hence improve the performance of rating prediction. Following these works, we modify the rating predictor by adding both user and item biases as follows:
\begin{equation}
\hat{y}_{u,i} = \hat{y}_{u,i}(\mathbf{z}_{u,i}) + b_u + b_i\label{eqn:fmwithbias}
\end{equation}
where $b_u$ and $b_i$ are the corresponding bias for user $u$ and item $i$, respectively. $\hat{y}_{u,i}$ is the predicted rating. We take the square loss as the objective function for parameter optimization.
\begin{equation}
J_{sqr} = \sum_{(u,i) \in O} (y_{u,i} - \hat{y}_{u,i})^2 + \lambda_{\Theta} \|\Theta\|^2 \label{eqn:sqrloss}
\end{equation}
where $O$ denotes the set of observed user-item rating pairs, $y_{u,i}$ is the observed rating score for user $u$ on item $i$,
$\Theta$ denotes all the parameters. The second term of Equation~\ref{eqn:sqrloss} is used as regularization to prevent the model from overfitting.

\subsection{Review-based Feature Learning} \label{ssec:review}
\begin{figure}
\centering
\includegraphics[width=.98\linewidth]{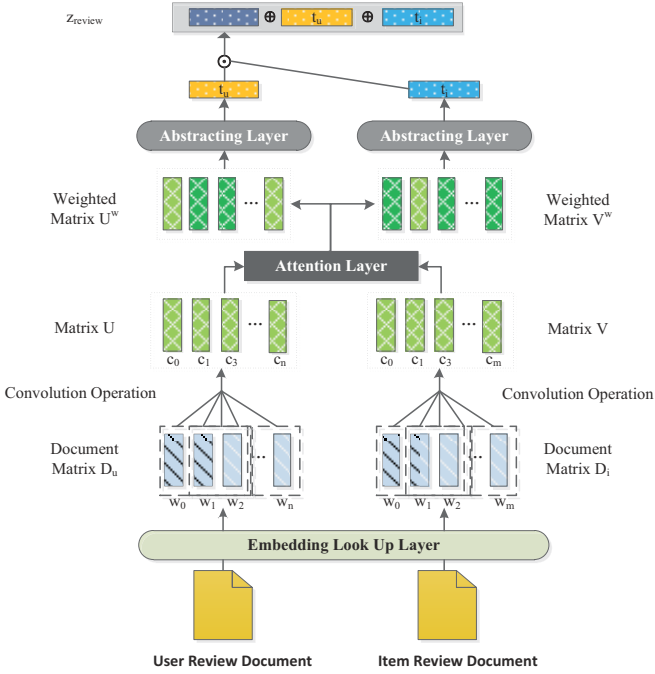}
\caption{The network architecture of the review-based feature learning.}\label{fig:review}
\end{figure}

Since the reviews made by a user reflect her preference, we take all the reviews written by the same user to form a single document as user review document. Similarly, we merge all the reviews made by the users for an item as item review document. It is expected that the semantic information covered in two kinds of review documents are quite different. While a user review document would contain more personal preference, an item review document is mainly comprised of different aspects focused on by all the relevant users. The task of review-based feature learning in \FCAN is to infer a latent feature vector for each user-item pair by jointly considering their review documents\footnote{When both user and item review documents are applicable, we use \textit{review document} instead for simplicity.}. The convolution operation has been successfully adopted in many natural language processing and information retrieval tasks like document representation learning~\cite{jmlr11:nlp,acl14:cnn1,emnlp14:cnn2,nips:lihangcnn}. Specifically, we use the convolution operation to extract different aspects covered by the review documents. Then we utilize an attentive layer to highlight the relevant aspects by jointly considering both the user preference and the characteristics of the item. At last, an abstracting layer is utilized to derive the final latent feature vector for the user-item pair. Figure~\ref{fig:review} demonstrates the network architecture for the review-based feature learning.

\paratitle{Convolution Layer.}
Given a review document $D = (w_1, w_2, ..., w_n)$, an embedding look-up layer first projects each word to its corresponding embedding $\mathbf{w}_i$ $\in\mathbb{R}^{1\times t}$. Then a document matrix $\mathbf{D}$ is formed by concatenating these embeddings in the order of their appearances in the document:
  $$\textbf{D} = \left[ \begin{matrix}
	\cdots \textbf{w}_{j-1}, \textbf{w}_j, \textbf{w}_{j+1} \cdots
  \end{matrix}\right]^\mathrm{T}$$
where $\mathbf{w}_j$ is word embedding of the word at $j$-th position in document $D$. That is, the order of words is preserved in matrix $\textbf{D}$, and in turn enables convolution layer to extract more accurate semantic information comparing to bag-of-words techniques~\cite{jmlr11:nlp}. Specifically, a convolution filter $f_j$ over a sliding window of size $s$ is used to extract the contextual feature $c_h^j$ from the local context. To be specific:
\begin{equation}
c_h^j = \sigma(\mathbf{\textbf{W}}_c^j\textbf{D}_{h:h+s-1})\label{eqn:cnn}
\end{equation}
where $\sigma$ is the nonlinear activation function, $\mathbf{W}_c^j$ is the convolution weight vector for filter $j$, $\textbf{D}_{h:h+s-1}$ is the slice of matrix $\textbf{D}$ within the sliding window starting at $h$-th position. Without explicit specification, we opt for Rectified Linear Unit (ReLU) as the activation function, \ie $ReLU(x)=\max(x,0)$. Here, we pad $s-1$ zero vectors at the end of document matrix $\mathbf{D}$ to produce $n$ contextual features, where $n$ is the length of document $D$. We use multiple convolution filters with different convolution weight vectors to extract the contextual features for each word with its local context (\ie $s$ consecutive words). In this work, we use two different $\mathbf{W}_c^*$ for processing the user review document and item review document respectively.

\paratitle{Attentive Layer.}
As stated above, we assume that a user review document could contain more personal and different preferences on different items. Likewise, an item review document might consist of different aspects focused on by different users. In other words, not all information contained in the review documents could be useful for inferring the rating score for a specific user-item pair. To effectively capture useful information, We employ an attention layer to the review documents of the corresponding user-item pair.

After applying the convolution operation for a review document, we can form a contextual feature vector $\mathbf{c}_h$ for the word at $h$-th position in the document:
\begin{align}
\mathbf{c}_h=[c_h^1,\cdots,c_h^f]\label{eqn:c}
\end{align}
where $c_h^j$ is the contextual feature calculated by convolution filter $j$ for the $h$-th word by using Equation~\ref{eqn:cnn}. $\mathbf{c}_h$ is thus the contextual feature vector of the $h$-th word after convolution layer. In this way, we can form two matrices for the user and item review documents respectively:
\begin{align}
\mathbf{U}&=[\mathbf{c}_1^u;\cdots;\mathbf{c}_n^u]\\
\mathbf{V}&=[\mathbf{c}_1^i;\cdots;\mathbf{c}_m^i]
\end{align}
where $\mathbf{c}_j^u$ and $\mathbf{c}_k^i$ are the contextual feature vectors based on Equation~\ref{eqn:c} for the $j$-th word and the $k$-th word in the user and item review documents respectively, $n$ and $m$ are the lengths of the user and item review documents respectively.

Inspired by the work in~\cite{corr16:attention,acl17:cane}, we utilize an attentive matrix, $\mathbf{T} \in\mathbb{R}^{f \times f}$, to derive the importance of each contextual feature vector for both $\mathbf{U}$ and $\mathbf{V}$. In detail, we project matrix $\textbf{U}$ and $\textbf{V}$ into the same latent space and calculate the pair-wise relatedness between each pair of contextual feature vectors $\mathbf{c}_j^u$ and $\mathbf{c}_k^i$ as follows:
\begin{equation}
\textbf{R}_{j,k} = \tanh(\mathbf{c}_j^{u\mathrm{T}}\mathbf{T}\mathbf{c}_k^i) \label{eqn:attentive}
\end{equation}
where $\textbf{R}_{j,k}$ is the relatedness between $\mathbf{c}_j^u$ and $\mathbf{c}_k^i$, $\tanh$ is the hyperbolic tangent function.

Based on Equation~\ref{eqn:attentive}, a row $\mathbf{R}_{j,\ast}$ contains the relatedness scores between the contextual feature vector $\mathbf{c}_j^u$ in $\mathbf{U}$ and all the contextual feature vectors in $\mathbf{V}$. Similarly, a column $\mathbf{R}_{\ast,k}$ contains the relatedness scores between the contextual feature vector $\mathbf{c}_k^i$ in $\mathbf{V}$ and all the contextual feature vectors in $\mathbf{U}$. A mean-pooling operation is then applied to each row/column of $\mathbf{R}$ as follows:
\begin{align}
g_j^u &=\text{\textbf{mean}}(R_{j,1},\cdots,R_{j,m})\label{eqn:g_j}\\
g_k^i &=\text{\textbf{mean}}(R_{1,k},\cdots,R_{n,k})\label{eqn:g_k}
\end{align}

Based on the mean relatedness calculated in Equations~\ref{eqn:g_j} and~\ref{eqn:g_k}, we can highlight the importance of each contextual feature vector in $\mathbf{U}$ and $\mathbf{V}$ respectively:
\begin{align}
a^u_j &= \frac{\exp(g^u_j)}{\sum^n_{h} \exp(g^u_h)} \label{eqn:auh}\\
a^i_k &= \frac{\exp(g^i_k)}{\sum^m_{h} \exp(g^i_h)}
\end{align}
where $a^u_j$ and $a^i_k$ are the attentive weights of $\mathbf{U}_{j,\ast}$ and $\mathbf{V}_{k,\ast}$ respectively. At last, we obtain the attentive weight vectors $\textbf{a}^u$ and $\textbf{a}^v$ through the attentive layer.
\begin{align}
\mathbf{a}^u &=[a^u_1,\cdots,a^u_n] \label{eqn:uatt}\\
\mathbf{a}^i &=[a^i_1,\cdots,a^i_m] \label{eqn:iatt}
\end{align}

Here, $\mathbf{a}^u$ and $\mathbf{a}^i$ can be regarded as the learnt distribution of the degree of the importance to the words in user review document and item review document, respectively.

\paratitle{Abstracting Layer.}
We obtain a weighted $\mathbf{U}$ and $\mathbf{V}$ based on the attentive vectors $\mathbf{a}^u$ and $\mathbf{a}^i$ as follows:
\begin{align}
\textbf{U}^w &= \text{diag}(\textbf{a}^u)\textbf{U}\\
\textbf{V}^w &= \text{diag}(\textbf{a}^v)\textbf{V}\label{eqn:Vw}
\end{align}
where $\text{diag}(\textbf{a}^{\ast})$ is the diagonal matrix whose diagonal are elements in vector $\textbf{a}^{\ast}$. Recall that the attentive weights are calculated based on both the user review document and the item review document, a large weight indicates the corresponding contextual feature vector is more relevant to the user-item pair. In this sense, we can consider these highly weighted contextual vectors as relevant aspects covered in the corresponding review documents for the user-item pair.

At this step, we can simply sum up the weighted contextual feature vectors to represent the user/item under consideration, by following the works in~\cite{corr16:attention,acl17:cane}. However, a simple weighted average could introduce too much noisy information since irrelevant aspects covered in both user and item review documents could account for a major proportion. Here, we choose to stack further neural transformations to extract higher-level semantic features based on $\textbf{U}^w$ and $\textbf{V}^w$ respectively. To accommodate with the noise and irrelevant aspects extracted above, we employ a mean-pooling CNN network to further abstract higher-level features $h^u_a$ as follows:
\begin{equation}
\begin{aligned}
h^j_{h} &= \sigma(\mathbf{W}^j_{a}\mathbf{U}^w_{h:h+s-1})\\
h_j &= \text{\textbf{mean}}(h^j_1, \cdots, h^j_{n}) \\
\textbf{h}^u &= [h_1, \cdots, h_f]
\end{aligned} \label{eqn:cnn2}
\end{equation}
where $\mathbf{W}^j_a$ is the convolution weight vector for filter $j$, and a sliding window of size $s$ is used. Similar process is applied to extract higher-level feature vector $\mathbf{h}^i$ from $\textbf{V}^w$. There is one straightforward merit for applying a CNN network over $\textbf{U}^w$ and $\textbf{V}^w$. Note that all the contextual feature vectors in $\textbf{U}$ and $\textbf{V}$ are extracted based on the convolution operation with a local context window. A further convolution operation inside the CNN network could cover a larger context for latent feature extraction, but with relevance weighted information (\ie less noisy information). In detail, with a window size of $s$, Equation~\ref{eqn:cnn2} actually takes $2\cdot s-1$ words into consideration. And these words are weighted based on their relevance to the user-item pair, leading to a more precise higher-level semantic information extraction. The intuition behind our mean-pooling setting is that users could express their opinions on various aspects for an item in their reviews. For example, a movie fan considers not only the cast but also the director and the genre to make a rating. By using mean-pooling strategy, more relevant latent features would be extracted in abstraction layer. Instead, max-pooling strategy may ignore some important features due to its downsampling property such that only the most important feature is retained. Our experimental results also validate the superiority of using mean-pooling strategy (ref. Section~\ref{ssec:analysisreview}).

Note that we use different $\mathbf{W}_a^*$ for the users and items respectively in the CNN network. The purpose is to allow the two independent CNN models to project both $\mathbf{U}^w$ and $\mathbf{V}^w$ into the same latent space. At last, we stack one shared MLP layer to further extract higher-level features:
\begin{align}
\mathbf{t}_{u} &= \sigma(\mathbf{W}_1\mathbf{h}^u + \mathbf{b}_1) \label{eqn:mlp}\\
\mathbf{t}_{i} &= \sigma(\mathbf{W}_1\mathbf{h}^i + \mathbf{b}_1)
\end{align}
where $\mathbf{W}_1\in\mathbb{R}^{l \times f}$ is the transformation matrix and $\mathbf{b}_1$ is the bias vector.  Finally, we form the context-aware latent feature vector for the user-item pair as follows:
\begin{align}
\mathbf{z}_{\text{review}} &= [\mathbf{e}^{\text{review}}_{u,i} \oplus \mathbf{t}_u \oplus \mathbf{t}_i]\label{eqn:reviewvector}\\
\mathbf{e}^{\text{review}}_{u,i} &=\mathbf{t}_u \odot \mathbf{t}_i\label{eqn:reviewodot}
\end{align}
where $\odot$ is the element-wise product of vectors, $\oplus$ is the vector concatenation operation. Recall that $\mathbf{t}_u$ and $\mathbf{t}_i$ are dynamic and dependent on the user and item jointly (Equations~\ref{eqn:attentive}-\ref{eqn:Vw}). The element-wise product $\odot$ used for $\mathbf{e}^{\text{review}}_{u,i}$ further enhances the latent feature interactions. Hence, the derived latent feature vector $\mathbf{z}_{\text{review}}$ for the user-item pair captures both the individual characteristics and their interactions together. \FCAN then takes $\mathbf{z}_{\text{review}}$ through a linear combination and higher-order interaction modeling by using FM for rating prediction (ref. Equation~\ref{eqn:fmwithbias}), denoted as $\hat{y}_{u,i}(\mathbf{z}_{\text{review}})$.

\subsection{Interaction-based Feature Learning}\label{ssec:interaction}
Although the textual reviews provide rich information about the user preferences and the characteristics of the items, the latent features $\mathbf{z}_{\text{review}}$ learnt above is just review-driven, and hence do not represent the users' rating behaviors to its fullness. Therefore, we devise a learning process for latent feature extraction by using the user-item rating scores.

\begin{figure}
\centering
\includegraphics[width=.70\linewidth]{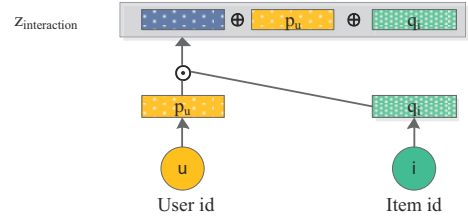}
\caption{The network architecture of the interaction-based learning.}\label{fig:interaction}
\end{figure}

We use a separate set of latent vectors for the users and items in interaction-based feature learning. Given the one-hot encoding of user/item identity $\textbf{x}_u/\textbf{x}_i$, we project it to its corresponding latent vector $\textbf{p}_u/\textbf{q}_i$ as
\begin{align}
\textbf{p}_u &=  \textbf{P}\textbf{x}_u\label{eqn:intrctnproju}\\
\textbf{q}_i &=  \textbf{Q}\textbf{x}_i\label{eqn:intrctnproji}
\end{align}
where $\textbf{P} \in\mathbb{R}^{l \times p}$ and $\textbf{Q} \in\mathbb{R}^{l \times q}$ denote the latent feature matrices of users and items respectively\footnote{For simplicity, we restrict the dimension size of user and item latent feature vectors to be identical in two learning components.}, $p$ and $q$ are the numbers of users and items. Since no context information like textual reviews is available, we then use an element-wise product operation to extract interaction based features.
\begin{align}
\textbf{e}^{\text{int}}_{u,i} = \mathbf{p}_u \odot \mathbf{q}_.\label{eqn:intodot}
\end{align}
Similar to Equation~\ref{eqn:reviewvector}, we form the context-awre latent feature vector for the user-item pair as follows:
\begin{align}
\mathbf{z}_{\text{interaction}} = [\mathbf{e}^{\text{int}}_{u,i} \oplus \mathbf{p}_u \oplus \mathbf{q}_i]\label{eqn:interaction}
\end{align}
Figure~\ref{fig:interaction} depicts the interaction-based learning process on the basis of user-item pairs. We feed $\mathbf{z}_{\text{interaction}}$ as input into FM for rating score predication, denoted as $\hat{y}_{u,i}(\mathbf{z}_{\text{interaction}})$.

\subsection{Fusion}
The two learning components described above extract the context-aware latent features from two different information sources. It is expected that the integration of the two components could complement each other and yield better prediction performance. A simple solution is to linearly interpolate the estimated rating score as follows:
\begin{align}
\hat{y}_{u,i} = \alpha \hat{y}_{u,i}(\mathbf{z}_{\text{review}}) + (1 - \alpha)\hat{y}_{u,i}(\mathbf{z}_{\text{interaction}}) + b_u + b_i \label{eqn:simplefusion}
\end{align}
where parameter $\alpha$ works as a tradeoff between the two components. However, linearly fusing the two models with a static $\alpha$ may not be a suitable choice for rating prediction. It is likely that a user could give a high score for an item because of some specific preferred features, but ignore the other moderate characteristics. Analogously, we introduce a dynamic weighting scheme by preferring the component with the higher rating prediction (\ie either the knowledge from the explicit reviews or other factors). Specifically, we calculate $\alpha$ as follows:
\begin{align}
\alpha = \frac{\hat{y}_{u,i}(\mathbf{z}_{\text{review}})}{ \hat{y}_{u,i}(\mathbf{z}_{\text{review}})+\hat{y}_{u,i}(\mathbf{z}_{\text{interaction}})}\label{eqn:dynamic}
\end{align}
In Equation~\ref{eqn:dynamic}, parameter $\alpha$ becomes larger when the review-based component predicates a higher score than the interaction-based component does. 

\subsection{Optimization of Model}

The parameters of \FCAN are optimized based on Equations~\ref{eqn:sqrloss} and~\ref{eqn:simplefusion} with stochastic gradient descent (SGD) and back-propagation. That is, the parameters for both two learning components are jointly learnt. For parameter update, we utilize RMSprop~\cite{lecture12rmsp} over mini-batches. Additionally, to prevent overfitting, we adopt dropout~\cite{jmlr14:dropout} strategy to the MLP layers.

\section{Experiment}\label{sec:exp}
In this section, we conduct extensive experiments on five real-world datasets for performance evaluation. We also analyze the contributions of the two components and different settings for \FCAN\footnote{The implementation will be released after paper acceptance.}. Finally, a thorough analysis of review-based feature learning and a case study are presented.

\subsection{Experimental Settings}

\paratitle{Datasets.}  Five Amazon $5$-cores datasets\footnote{http://jmcauley.ucsd.edu/data/amazon/}~\cite{www16:datasets} are used for performance evaluation: \textit{Musical Instruments}, \textit{Office Products}, \textit{Digital Music}, \textit{Video Games} and \textit{Tools Improvement}. These datasets consist of users' explicit ratings on items ranging from $1$ to $5$, and contain the textual reviews made by the users. Following the preprocessing steps used in~\cite{recsys16:convMf}, we perform the preprocessing for the review documents for all datasets as follows: 1) remove stop words and words that have the document frequency higher than $0.5$; 2) calculate tf-idf score for each word and select the top $20,000$ distinct words as vocabulary; 3) remove all words out of the vocabulary from raw documents; 4) amputate (pad) the long (short) review documents to the same length of $300$ words. We further filter out the rating records which contain empty review after document preprocessing.
\begin{table*}
\centering
\caption{Statistics of the five datasets}
  \begin{tabular}{cccccccc}
  \toprule
  Datasets & \# users & \# items & \# ratings & \# words per review & \# words per user & \# words per item & density\\
  \midrule
  Musical Instruments & $1,429$ & $900$ & $10,261$ & $32.45$ & $141.32$ & $200.12$ & $0.798\%$\\
 Office Products & $4,905$ & $2,420$ & $53,228$ & $48.15$ & $197.93$ & $229.52$ & $0.448\%$ \\
  Digital Music & $5,540$ & $3,568$ & $64,666$ & $69.57$ & $216.21$ & $266.51$ & $0.327\%$\\
  Video Games & $24,303$ & $10,672$ & $231,577$ & $72.13$ & $188.79$ & $260.60$ & $0.089\%$\\
  Tools Improvement & $16,638$ & $10,217$ & $134,345$ & $38.75$ & $162.53$ & $212.48$ & $0.079\%$\\
  \bottomrule \label{tab:stats}
  \end{tabular}
\end{table*}

\begin{table*}
\scriptsize
\centering
\caption{Overall performance comparison on three datasets. The best and second best results are highlighted in boldface and underlined respectively. $\blacktriangle\%$ denotes the improvement of \FCAN over the best baseline performer. $\dagger$ indicates that the difference to the best result is statistically significant at $0.01$ level.}
\begin{tabular}{c||c|c|c|c|c}
\toprule
Method & Musical Instruments & Office Products & Digital Music & Video Games & Tools Improvement\\
\midrule
PMF & $1.401^\dagger$ & $1.091^\dagger$ & $1.211^\dagger$ & $1.669^\dagger$ & $1.564^\dagger$ \\
\hline
CDL & $0.861^\dagger$ & $\underline{0.754}^\dagger$ & $0.882^\dagger$ & $1.179^\dagger$ & $1.033^\dagger$ \\
\hline
RBLT & $0.815^\dagger$ & $0.757^\dagger$ & $\underline{0.872}^\dagger$ & $1.147^\dagger$ & $\underline{0.983}^\dagger$ \\
\hline
CMLE & $0.818^\dagger$ & $0.761^\dagger$ & $0.883^\dagger$ & $1.254^\dagger$ & $1.023^\dagger$ \\
\hline
ConvMF & $0.991^\dagger$ & $0.960^\dagger$ & $1.084^\dagger$ & $1.449^\dagger$ & $1.240^\dagger$ \\
\hline
DeepCoNN & $0.814^\dagger$  & $0.860^\dagger$ & $1.060^\dagger$ & $1.238^\dagger$ & $1.063^\dagger$ \\
\hline
TransNets & $\underline{0.799}^\dagger$  & $0.760^\dagger$ & $0.910^\dagger$ & $1.196^\dagger$ & $1.008^\dagger$ \\
\hline
D-attn & $0.984^\dagger$  & $0.824^\dagger$ & $0.914^\dagger$ & $\underline{1.142}^\dagger$ & $1.046^\dagger$ \\
\midrule
\FCAN & $\textbf{0.776}$ & $\textbf{0.722}$ & $\textbf{0.831}$ & $\textbf{1.065}$ & $\textbf{0.942}$ \\
\hline\hline
$\blacktriangle\%$ & $2.89\%$ & $4.24\%$ & $4.70\%$ & $6.74\%$ & $4.17\%$  \\
\bottomrule
\end{tabular} \label{tab:comparison}
\end{table*}

Table~\ref{tab:stats} summarizes the detailed statistics about the five datasets after preprocessing. We can see that the five datasets hold different characteristics. The Music Instruments dataset is the smallest, but its interaction matrix is the densest among all the five datasets. In contrast, Video Games and Tools Improvement are the two largest datasets and are much sparser. For evaluation, we randomly select $80\%$ of each dataset as the training set and the remaining $20\%$ as the testing set. We further split $10\%$ of the training set as the validation set for hyper-parameter validation. The training sets are selected such that at least one interaction for each user and item should be included. Following the work in~\cite{recsys17:transnet}, the reviews in the validation set and testing set are excluded since they are unavailable during rating prediction.

\paratitle{Baselines}
We compare the proposed \FCAN against the following state-of-the-art rating prediction methods:
\begin{itemize}
\item[$\bullet$] PMF: Probabilistic matrix factorization is a standard matrix factorization model that uses only rating scores~\cite{nips07:pmf}. We use the Alternating Least Squares (ALS) techniques for model optimization.
\item[$\bullet$] CDL: Collaborative Deep Learning~\cite{kdd15:cdl} is the first heirarchical Bayesian model to build the connection between deep learning technique (SDAE)~\cite{jmlr10:sdae} and matrix factorization model. Following the adaption used in~\cite{recsys16:convMf}, we set the confidence parameter to $1$ if the rating is observed and $0$ otherwise.
\item[$\bullet$] RBLT: Rating-Boosted Latent Topics model integrates both  matrix factorization model and topic model~\cite{ijcai16:rblt}. It proposes a rating-boosted approach which utilizes the rating-boosted reviews and rating scores together for rating prediction.
\item[$\bullet$] CMLE: Collaborative Multi-Level Embedding Learning integrates word embedding model with matrix factorization to learn user and item embeddings~\cite{ijcai16:cmle}. Given a new user-item pair, the rating can be predicted by the dot-product of its user and item embeddings.
\item[$\bullet$] ConvMF: Convolutional Matrix Factorization integrates CNN into PMF for rating prediction~\cite{recsys16:convMf}. The item latent features are extracted by using CNN over the item review documents.
\item[$\bullet$] DeepCoNN: Deep Cooperative Neural Networks uses two parallel CNN networks to extract latent feature vectors from both the user review documents and item review documents~\cite{wsdm17:deepconn}. FM is then used for rating prediction.
\item[$\bullet$] D-attn: Dual local and global attention model leverages global and local attentions to enable an interpretable embedding of users and items~\cite{recsys17:dattn}. Finally, the rating can be estimated by dot-product of the user and item embeddings.
\item[$\bullet$] TransNets: TransNets extends the DeepCoNN model by adding an additional layer to represent the target user-item review, which is unavailable at test time~\cite{recsys17:transnet}. Then TransNets can mimic the target user-item review representation at test time and thus improve the performance of rating prediction.
\end{itemize}
The first method listed above is the conventional latent model that utilizes only the user-item rating scores. The rest are the methods that utilize the review documents for rating prediction. D-attn is an attention-based recommendation model.

\paratitle{Hyper-parameter Settings}
We use grid search to tune the hyper-parameters for all the methods based on the setting strategies reported by their papers, and report their best performances over 5 runs on the testing set. The latent dimension is optimized from $\{15, 25,50,100,150,200,300\}$. The embedding dimension of words in all models is set to $300$. The batch size for Musical Instruments, Office Products and Digital Music is set to $100$. For the other two big dataset (Video Games and Tools Improvement), the batch size is set to $200$. The number of convolution filters is set to $50$. The statistical significance is conducted by applying the student \textit{t-test}.

For \FCAN, the dimension for user and item latent feature vectors is $l = 15$, window size is $s = 3$, and $v$ is set to $50$ for FM rating prediction. The dropout rate is set to $0.2$. And the learning rate is set to $0.001$. The regularization parameter $\lambda_{\Theta}$ is tuned  from $[0.05, 0.01, 0.005, 0.001]$ for the five datasets.

\paratitle{Evaluation Metric}
The well-known Mean Square Error (MSE) is adopted for performance evaluation:
\begin{equation}
MSE = \frac{1}{|O_t|}\sum_{(u,i)\in O_t}(y_{u,i} - \hat{y}_{u,i})^2 \label{eqn:mse}
\end{equation}
where $O_t$ is the set of the user-item pairs in the testing set.

\subsection{Performance Evaluation}

The overall performances of all the methods are reported in Table~\ref{tab:comparison}. The best and the second best results are highlighted in boldface and underlined respectively. Here, we make the following observations.

First, we can see that PMF performs the worst in all five datasets. Among the five datasets, PMF performs much worse on Video Games and Tools Improvement than on the other three ones. It is reasonable since these two datasets have the sparsest user-item interaction data. All review-based rating prediction methods evaluated here perform much better than PMF. This proves that incorporating review information can provide more semantic information for understanding the user rating behaviors than using a single rating score.

Second, among review-based baseline methods, RBLT performs relatively good across the five datasets. By leveraging the user and item biases, RBLT successfully achieves the second best on Digital Music and Tools Improvement datasets. Also, RBLT obtains very close performance to the best baseline performer on the other three datasets. Such good performances should be attributed to its rating-boost reviews, which ensure the preferred features in high-rating reviews to be extracted successfully. On the other hand, ConvMF and DeepCoNN achieve varying and unstable performance across the five datasets. For the datasets with long reviews (i.e., Office Product, Digital Music and Video Games), they obtain relatively poorer performance, which indicates that both ConvMF and DeepCoNN fail to extract relevant features from long reviews and the prediction performance are adversely affected by the noise and irrelevant information within the reviews. The extended DeepCoNN method (TransNets) outperforms DeepCoNN across five datasets, which is consistent with the results reported in~\cite{recsys17:transnet}. D-attn achieves marginal improvement over ConvMF on three datasets. However, on Digital Music and Video Games, D-attn obtains siginificant improvement over both DeepCoNN and ConvMF, which demonstrates that attention mechanism does help CNN capture relevant information from reviews. This observation is also in line with the experimental results reported in~\cite{recsys17:dattn},

Third, \FCAN consistently achieves the best MSE scores across the five datasets. The last row indicates the relative improvements of \FCAN over the best baseline performer. We can observe that the improvements gained by \FCAN are consistent and stable. On average, the relative improvement of \FCAN against the best baseline is $4.55\%$. Even for the sparsest dataset\textemdash Tools Improvement, it still gains $4.17\%$ improvement compared to the best baseline. This result implies that \FCAN is effective for rating prediction on datasets with different characteristics. Moreover, the significant performance gap between \FCAN and D-attn validates that the context-aware user-item representation learning devised for \FCAN captures more knowledge about users' rating behaviors by considering both the individial characteristics and their interactions.

\begin{table*}
\scriptsize
\centering
\caption{The impact of the two components in \FCAN. Review: review-based feature learning. Rating: interaction-based feature learning. Review-int: review-based feature learning without $\mathbf{e}^{\text{review}}_{u,i}$. Rating-int: interaction-based feature learning without $\mathbf{e}^{\text{int}}_{u,i}$. \textsc{CARL}+LR: a linear regression is utilized as the rating predictor. The best results are highlighted in boldface. }
\begin{tabular}{c||c|c|c|c|c}
\hline
Methods & Musical Instruments & Office Products & Digital Music & Video Games & Tools Improvement\\
\hline
Rating-int & $0.796$ & $0.753$ & $0.955$ & $1.280$ & $1.017$ \\
\hline
Review-int & $0.783$ & $0.745$ & $0.885$ & $1.080$ & $0.961$ \\
\hline
Rating & $0.785$  & $0.744$ & $0.938$ & $1.270$ & $1.013$ \\
\hline
Review & $0.782$ & $0.740$ & $0.862$ & $1.087$ & $0.955$\\
\hline\hline
\FCAN & $\textbf{0.776}$ & $0.722$ & $\textbf{0.831}$ & $\textbf{1.065}$ & $\textbf{0.942}$ \\
\hline
\textsc{CARL}+LR & $0.779$ & $\textbf{0.714}$ & $0.842$ & $1.069$ & $0.944$ \\
\hline
\end{tabular} \label{tbl:cmpfus}
\end{table*}

\subsection{Analysis of \FCAN}
\label{ssec:FCANAnalysis}
We now study the impact of different model settings for \FCAN.

\begin{figure}
\centering
\includegraphics[width=.65\columnwidth]{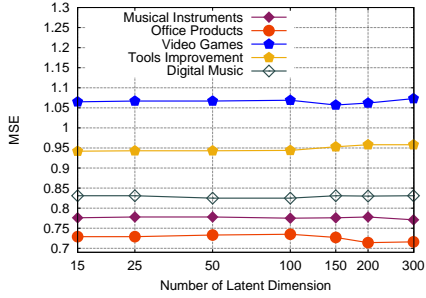}
\caption{Impact of dimension number $l$ across the five datasets.}\label{fig:dim}
\end{figure}

\paratitle{Number of Dimensions.}
Figure~\ref{fig:dim} plots the performance of \FCAN by varying the latent dimension size $l$ in $\{15,25,50,100,150,200,300\}$. We can see that \FCAN performs consistently well across a wide range of $l$ values (\ie $[15,300]$) with little performance variation. Even with a much smaller $l$ value (\ie $l=15$), \FCAN achieves nearly optimal rating prediction accuracy in most datasets. We believe this is attributed to the context-aware feature learning on the basis of the user-item pair. Although a much larger $l$ (\eg $l\geq 150$) could further obtain a bit performance gain for Office Products and Video Games, the resultant computation cost is much larger because FM utilizes the second-order feature interactions for rating prediction. Accordingly, we use $l=15$ in the experiments.

\eat{
we can see that \FCAN achieves the  rating accuracy when the latent dimension is small (\ie $l=15$) in most datasets. It demonstrates that \FCAN can represent the user and item well with a small number of latent features. We believe this is attributed to the review-based feature learning on the basis of the user-item pair utilized in \FCAN. Accordingly, we use $l=15$ in the experiments. Second, \FCAN performs consistently well across a wide range (\ie $[15,300]$) with little variations. Notice that, for Office Products and Video Games, when $l$ is set relatively larger (\ie $l\geq 150$), the performance of \FCAN starts to improve. This is reasonable since there are long user review documents and item review documents in the two datasets (ref. Table~\ref{tab:stats}), a larger dimension could capture more number of meaningful context features and therefore better model the relationships between user and item.
}

\begin{figure}[t]
\centering
\begin{subfigure}[Video Games]{\includegraphics[width=.45\linewidth] {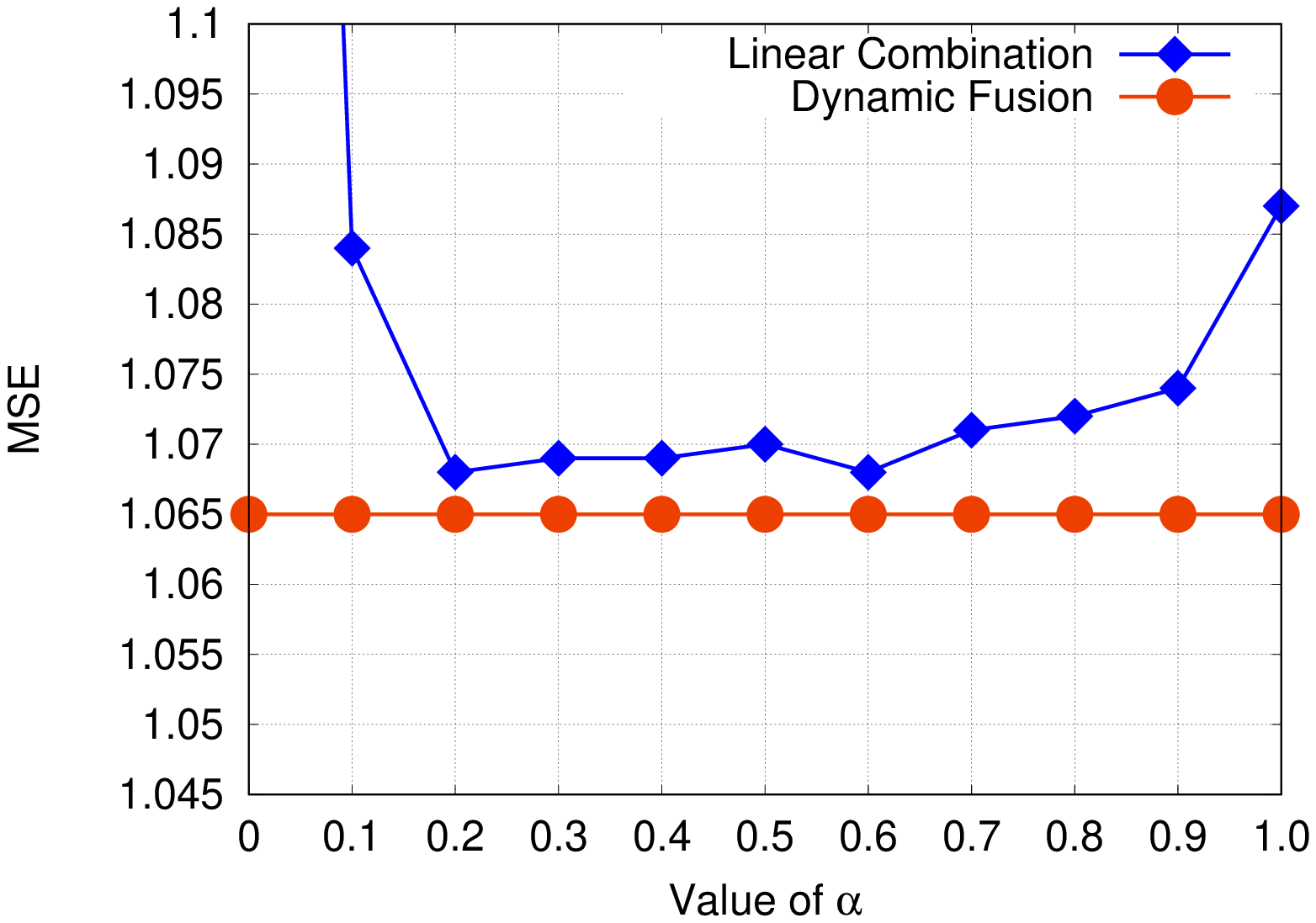}
   \label{fig:alpha:video}
 }%
\end{subfigure}\hfill
 \begin{subfigure}[Musical Instruments]{\includegraphics[width=.45\linewidth]{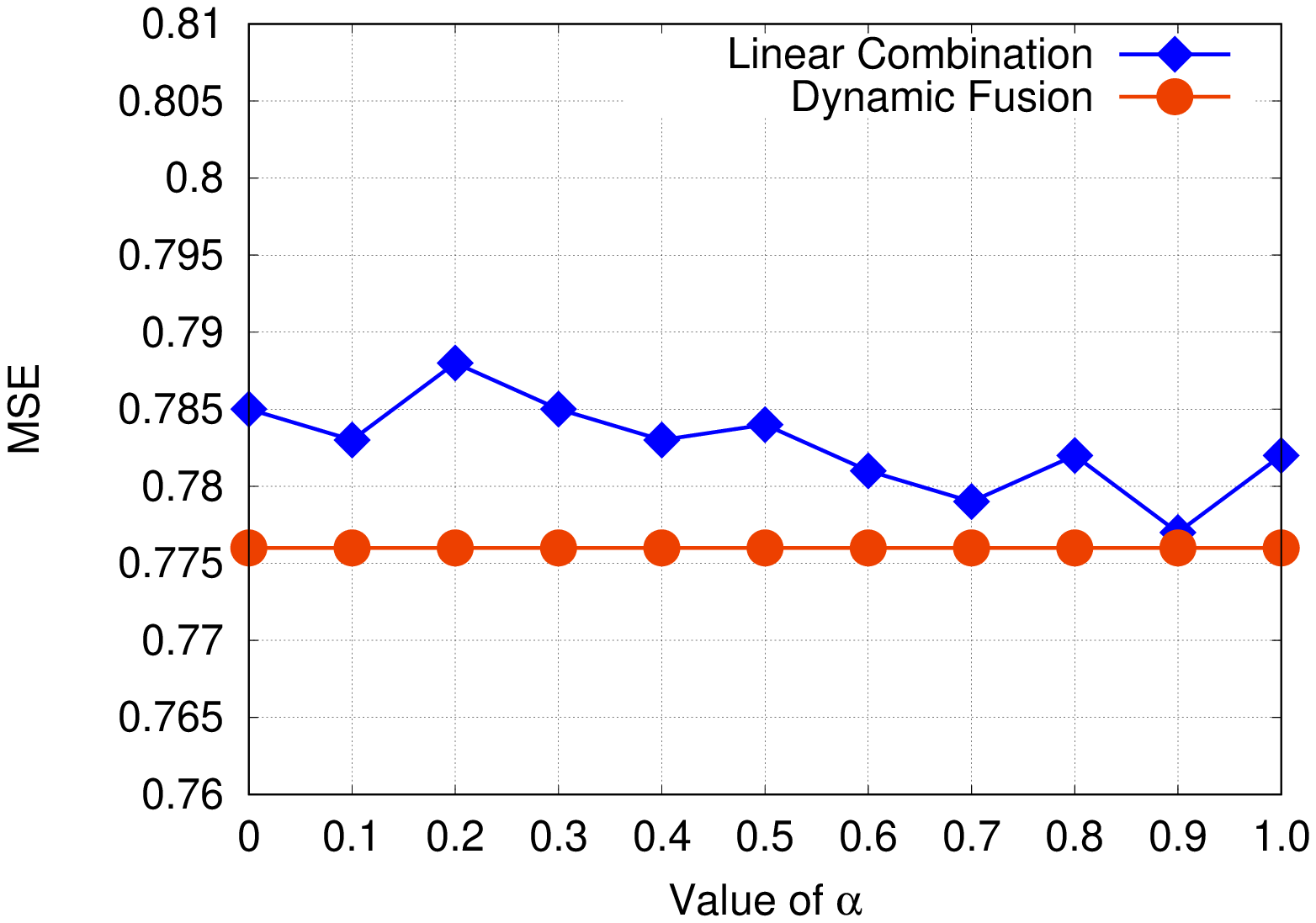}
   \label{fig:alpha:music}
 }%
\end{subfigure}
\caption{Impact of dynamic linear fusion on Video Games (a) and Musical Instruments (b).}
\end{figure}

\begin{table*}
\scriptsize
\centering
\caption{The impact of the attentive layer in the review-based feature learning. Review: review-based feature learning. Review-att: attentive layer is removed. The best and second best results are highlighted in boldface and underlined respectively. $\blacktriangle\%$ denotes the improvement of Review against Review-att.}
\begin{tabular}{c||c|c|c|c|c}
\hline
Methods & Musical Instruments & Office Products & Digital Music & Video Games & Tools Improvement\\
\hline
DeepCoNN & $0.814$  & $0.860$ & $1.060$ & $1.238$ & $1.063$ \\
\hline
Review-att & $\underline{0.813}$ & $\underline{0.766}$ & $\underline{0.933}$ & $\underline{1.108}$ & $\underline{1.028}$\\
\hline
Review & $\textbf{0.782}$ & $\textbf{0.740}$ & $\textbf{0.862}$ & $\textbf{1.087}$ & $\textbf{0.955}$ \\
\hline\hline
$\blacktriangle\%$ & $3.81\%$ & $3.39\%$ & $7.61\%$ & $1.89\%$ & $7.10\%$\\
\bottomrule
\end{tabular} \label{tbl:cmpfattn}
\end{table*}

\paratitle{The Impact of Two Learning Components.}
Recall that we use two independent learning components to derive the context-aware latent features from the reviews and user-item rating scores respectively. We further study the impact of each component (\ie review-based feature learning and interaction-based feature learning). It is worthwhile to note that parameter $l$ determines the model capacity. For fair comparison, we set the dimension size to be $l=30$ for each component, such that the final feature dimensions feeded into FM is equivalent to \FCAN with $l=15$ (ref. Equations~\ref{eqn:reviewvector} and~\ref{eqn:interaction}). Table~\ref{tbl:cmpfus} lists the prediction accuracy of the two components and \FCAN. We can see that, on the five datasets, the performance of the fusion model \FCAN is much better than the performances of the two components. In essence, \FCAN can reap the benefits from both reviews and user-item rating scores since the two independent components complement each other, and therefore the combination leads to better prediction accuracy.

Besides, we observe that both two components achieve comparable or even better performance than all state-of-the-art baseline methods (ref. Table~\ref{tab:comparison} and~\ref{tbl:cmpfus}). The interaction-based feature learning component (Rating) outperforms all baselines on the dense datesets (\ie Musical Instruments and Office Products). Meanwhile, as an interaction-based method, it is comparable to some review-based methods on the other three datasets. 

\eat{
Note that we derive a user-item pair based latent feature in the interaction-based feature learning component (see Equation~\ref{eqn:mlp1} and~\ref{eqn:interaction}). The experiments show these high-level and nonlinear features is beneficial for better rating prediction. This finding is also made in a recent interaction-based neural model~\cite{www17:ncf}.
}

Another observation is that using review-based feature learning (Review) alone obtains the best rating prediction accuracy against all baseline methods across the five datasets (ref. Table~\ref{tab:comparison} and~\ref{tbl:cmpfus}). Although in some cases, the prediction performance delivered by review-based feature learning is very close to the ones obtained by interaction-based feature learning. We need to emphasize that for sparse dataset, the review-based feature learning substantially outperforms the interaction-based ones. This observation is in line with the previous studies. Compared with ConvMF and DeepCoNN, the two CNN-based neural methods, review-based feature learning achieves significantly better rating prediction accuracy, though they all utilize the convolution operations to extract the semantic information from the reviews in the first place. Moreover, in contrast to D-attn, the attention-based neural method, review-based feature learning also gains large improvement. Overall, these observations confirm the superiority of context-aware user-item representation learning devised in \FCAN.

\paratitle{The Impact of Latent Feature Interactions}
The latent feature interaction is an important building block for \FCAN. Actually, we exploit different kinds of latent feature interactions for context-aware user-item representation learning. The user and item latent features $\mathbf{t}_u$ and $\mathbf{t}_i$ are drived by the review-based learning component based on the semantic interactions between the review documents. We further introduce $\mathbf{e}^{\text{review}}_{u,i}$ and $\mathbf{e}^{\text{int}}_{u,i}$ in the final user-item representation to boost the latent feature interactions. Also, we utilize the Factorization Machines (ref. Equation~\ref{eqn:fm}) to model the higher-order feature interactions for rating prediction. Here, we first examine the impact of $\mathbf{e}^{\text{review}}_{u,i}$ and $\mathbf{e}^{\text{int}}_{u,i}$ via an ablation test for two learning components respectively. We include the prediction performances by removing these two element-wise product based latent features from the two components (Review-int/Rating-int) respectively in Table~\ref{tbl:cmpfus}. We can see that both components experience performance degradation in almost all cases, except for Review-int on Video Games dataset. Another observation is that the interaction-based learning component achieves relatively worser performance on three samller but denser datasets (\ie Musical Instruments, Office Products and Digital Music). We believe that the interaction based knowledge learnt through $\mathbf{e}^{\text{review}}_{u,i}$ and $\mathbf{e}^{\text{int}}_{u,i}$ can capture more relevant information from much noisy data. 

We further examine the impact of using FM as rating predictor in \FCAN. Here, we replace the FM with a linear regression over the user-item representations (\ie $\mathbf{z}_{\text{review}}$ and $\mathbf{z}_{\text{interaction}}$) for rating prediction. The last row of Table~\ref{tbl:cmpfus} reports the prediction performance with this setting (\textsc{CARL}+LR).  Although \FCAN achieves better prediction accuracy with FM in most cases, we observe that the advantage is relatively weak except for Digital Music dataset. This is reasonable since the two learning components in \FCAN already model latent feature interactions explicitly from the reviews and rating scores respectively, the undiscovered knowledge regarding users' rating behaviors left for FM would be marginal for many cases. We also evaluate the variation by adding one layer of MLP with nonlinearity within \textsc{CARL}+LR. In detail, we utilize an one-layer MLP to further extract higher-level features from both $\mathbf{z}_{\text{review}}$ and $\mathbf{z}_{\text{interaction}}$ respectively. Then, the linear regression is used to calculate the prediction instead of using FM. However, further worse performance is experienced (results not shown).

\paratitle{The Impact of Dynamic Linear Fusion.}
Recall in Equation~\ref{eqn:dynamic}, we adopt a dynamic linear fusion mechanism to prefer the component with a higher rating prediction. Figure~\ref{fig:alpha:video} and~\ref{fig:alpha:music} plots the performance comparison between the dynamic fusion and the static linear combination by fixing $\alpha$ to a specific value in Equation~\ref{eqn:simplefusion}. We can see that the dynamic fusion achieves better prediction accuracy than the linear combination with different $\alpha$ settings. What's more, dynamic fusion can eliminate the need of parameter tuning required in the static linear combination. Similar observations are also made in the other three datasets.

\subsection{Analysis of Review-based Feature Learning}
\label{ssec:analysisreview}
To get a better understanding of the review-based learning component, we further analyze the impact of two key layers: attentive layer and abstracting layer.

\paratitle{The Impact of Attentive Layer.}
In review-based feature learning (Review), we adopt an attentive layer to capture relevant information upon the user-item pair. We evaluate the impact of attentive layer via an ablation test by removing attentive layer from the review-based feature learning component. To eliminate the impact of attentive layer, we set the attentive weights to be 1 in Equations~\ref{eqn:uatt} and~\ref{eqn:iatt} (Review-att). Table~\ref{tbl:cmpfattn} shows the performance comparison. We can see that when the attentive layer is applied, the performance of review-based component is significantly better than Review-att across the five datasets. The relative improvement is $4.76\%$ on average. The encouraging improvement confirms that the attentive layer is able to render the model to perform better prediction by identifying the relevant information for user-item pairs. A case study of attentive layer for prediction interpretation is later included in the next section.

\begin{figure}[t]
\centering
\begin{subfigure}[Tools Improvement]{\includegraphics[width=.45\linewidth] {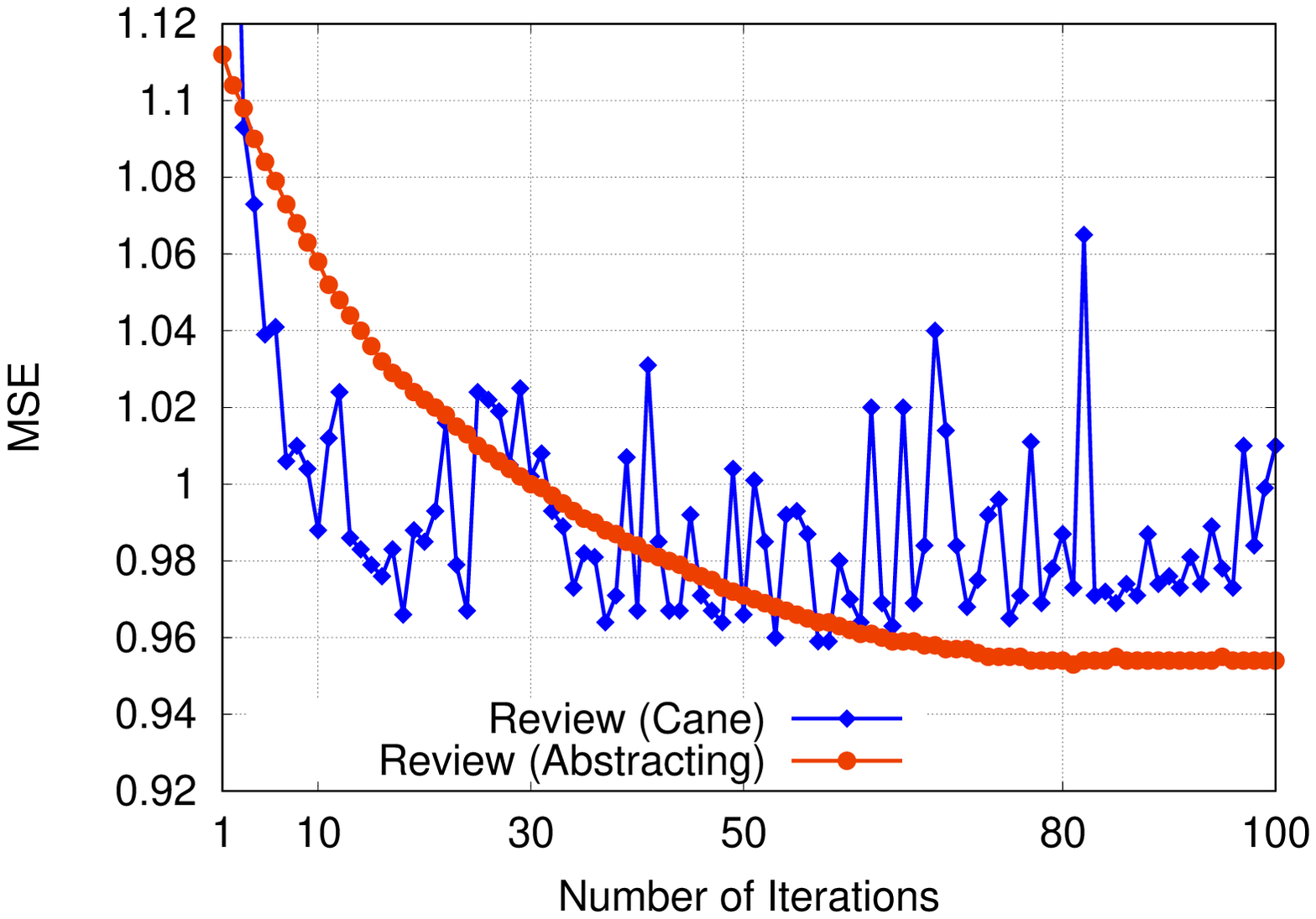}
   \label{fig:abs:video}
 }%
\end{subfigure}\hfill
 \begin{subfigure}[Musical Instruments]{\includegraphics[width=.45\linewidth]{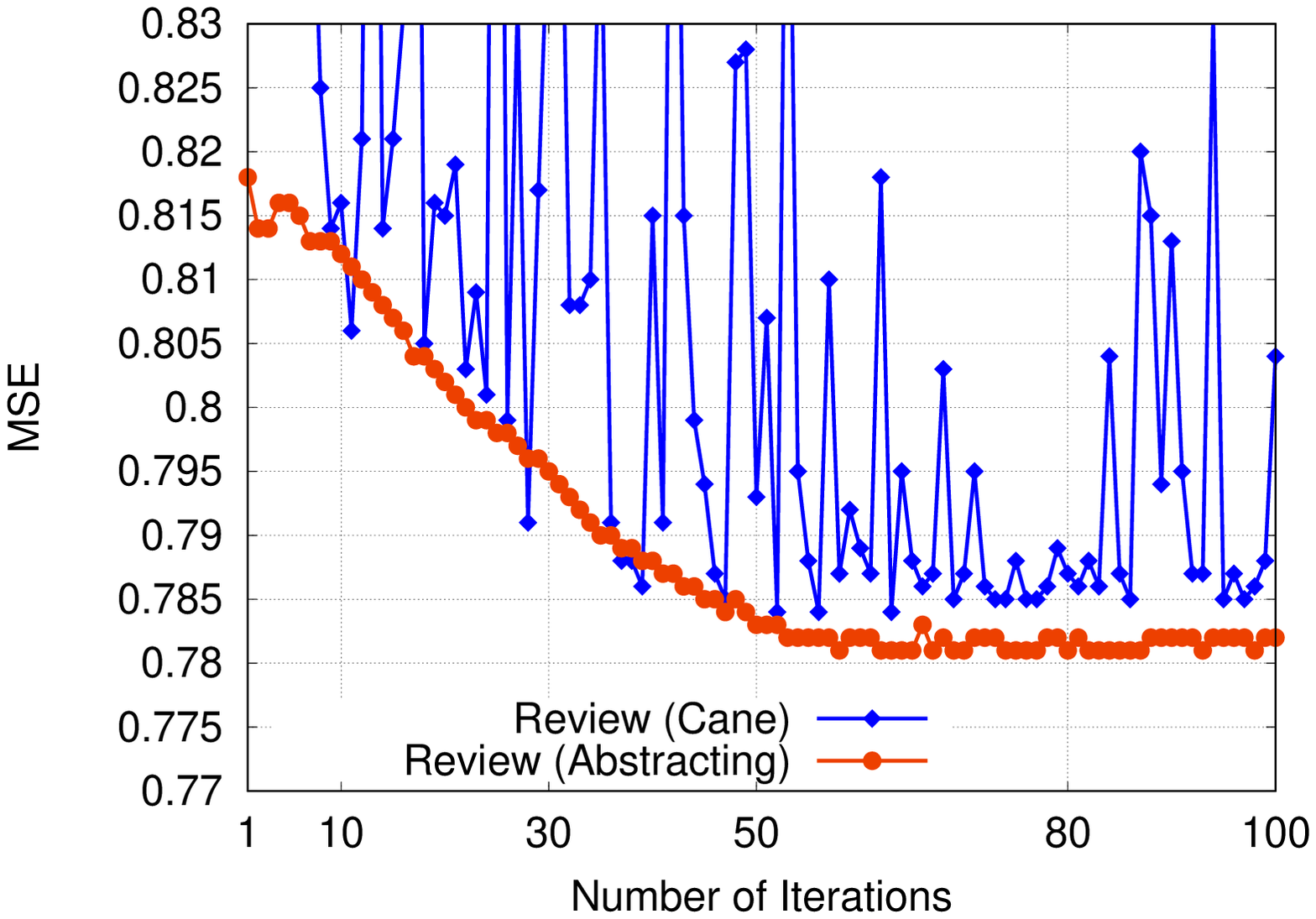}
   \label{fig:abs:music}
 }%
\end{subfigure}
\caption{Performance comparison per iteration between review-based feature learning with Abstracting layer and with CANE based strategy on Tools Improvement (a) and Musical Instruments (b).}
\end{figure}

\paratitle{The Impact of Abstracting Layer.}
Apart from the attentive layer, the abstracting layer is another pivotal layer in review-based feature learning. Recall that, Review-att has eliminated the impact of attentive layer and thus can be regarded as an extension of Deep-CoNN, which adds an abstracting layer after convolution operations. In Table~\ref{tbl:cmpfattn} we can observe that Review-att consistently outperforms DeepCoNN on the five datasets. In particular, for Office Products, Digital Music and Video Games datasets which contain relatively long review documents, Review-att gains substantial improvement over DeepCoNN. We attribute the improvement to the abstracting layer due to the fact that a larger context is considered for higher-level semantic information extraction via stacking the convolution operations.

In contrast to applying the abstracting layer,~\cite{acl17:cane} proposes a simple alternative by summing up the weighted contextual representations into one single vector as derived vertex embedding (named CANE). Though CANE has proven effective in modeling dynamic relations between vertices for network modeling, we have to point out that the textual information in~\cite{acl17:cane} is different from ours. In~\cite{acl17:cane}, the textual information is research papers written by the corresponding authors (\ie vertices) while the textual information we are dealing with is the concatenation of the reviews upon different user-item pair. Apparently, review documents are less coherent and more diverse than the research papers. Lots of noise are likely to be included by merely summing up the weighted contextual features. To tackle this issue, we employ the abstracting layer upon the attentive layer. Furthermore, we believe that an abstracting layer can extract higher-level semantic features from the weighted contextual features in a larger range. Figure~\ref{fig:abs:video} and~\ref{fig:abs:music} show the performance comparison per iteration between the review-based feature learning with abstracting layer and with CANE based alternative on Video Games and Musical Instruments, respectively. First, we can observe that the better prediction accuracy is achieved with the inclusion of the abstracting layer. Second, in terms of convergence, it is obvious that the performance by including the abstracting layer gradually becomes stable while CANE based strategy remains fluctuating within a large range. With the abstracting layer, review-based learning component almost reaches convergence after $50$ iterations on both datasets. These observations demonstrate that the abstracting layer is capable of extracting higher-level semantic features from the weighted contextual features and therefore enhancing the performance and robustness of the model.

In abstracting layer we choose to utilize the mean-pooling strategy instead of the prevalent max-pooling strategy. The reason lies in the assumption that we believe mean-pooling operation would retains more numbers of important features than the max-pooling operation does. The performance comparison between the review-based feature learning with mean-pooling strategy (Review-avg) and the review-based feature learning with max-pooling strategy (Review-max) is reported in Table~\ref{tbl:pool}. Note that although Review-max achieves the best performance on Office Products dataset, it underperforms Review-avg on all other datasets. Especially for the two sparsest datasets (Video Games and Tools Improvement), Review-avg gains substantial improvement over Review-max. These results validate that mean-pooling strategy extracts more useful features, compared to Review-max. Accordingly, we employ mean-pooling operation in abstracting layer.
 
\subsection{Visualization of Attentive Layer}
To further understand the review-based feature learning component of \FCAN, we manually check whether the attentive layer can identify relevant semantic information from both user and item review documents on the basis of the user-item pair. Hence, we sample a user $u$ and two user-item rating records associated with $u$ in the test set of Musical Instruments. The heat maps of the review documents for both sampled pairs are visualized in Table~\ref{tbl:case}.  The stronger the background color is, the larger the attentive weight of corresponding word is. A triple ($user_u$, $item_i$, $y_{u,i}$) denotes a user-item interaction such that user $u$ rates item $i$ with a score $y_{u,i}$. Note that the two items demonstrated in Table~\ref{tbl:case} receive quite different rating scores from the same user (\ie $5.0$ vs. $3.0$). Although there are many different aspects mentioned in the user review document, it is clear to see that the two attentive heat maps towards the two different items are quite differnt. For $item_1$, lots of applausive words, such as ``charm'', ``fun'', ``cool'', ``cheap'' and ``win'', are highlighted with larger attentive weights. In the contrary, less positive sentimental words are found to be highlighted in the user review document for $item_2$. From the highlighted information in the two heat maps, we can speculate that ``tiny shape'' and ``practicing outside'' aspects about $item_1$ are the focuses of user $u$, while ``noise level'' and ``power voltage'' aspects about $item_2$ could be more importance instead.

\begin{table*}
\scriptsize
\centering
\caption{The impact of the pooling strategy in review-based feature learning (Review).}
\begin{tabular}{c||c|c|c|c|c}
\hline
Methods & Musical Instruments & Office Products & Digital Music & Video Games & Tools Improvement\\
\hline
Review-max & $0.783$ & $\textbf{0.726}$ & $0.865$ & $1.096$ & $0.982$\\
\hline
Review-avg & $\textbf{0.782}$ & $0.740$ & $\textbf{0.862}$ & $\textbf{1.087}$ & $\textbf{0.955}$ \\
\hline
\end{tabular} \label{tbl:pool}
\end{table*}

A further close look at the heat maps of the two item review documents confirms a lot about the speculations made above. For $item_1$, it is not diffcult to see that some aspects such as ``save valuable space'', ``flexibility'', ``low profile'' are highlighted. The reasonable price, good quality, compact and low-profile design are found to be relevant information. From $item_2$'s review document, we can see that many aspects related to the battery part, expensive price and the construction of the product are highlighted. The observed correlations between the heat map of user review document and the heat map of item review document reveal that \FCAN is effective in capturing the relevant semantic information in reviews for a user-item pairs. 

Note that the real user review towards an item is excluded from both user and item review documents for rating prediction, since the user review is unavailable before her consumption. For the two user-item pairs studied above, we then list the corresponding real reviews provided in the original dataset in the last row of Table~\ref{tbl:case}. We manually highlight the opinions expressed by the user with red color. We can notice that the good quality, compact shape and convient design of $item_1$ are highly appreciated while the user has neutral sentiment towards item $2$ due to the annoying battery compartment. These groud-truth knowledge further validates that the context-aware user-item representatino learning devised in \FCAN enables a better understanding about users' rating behaviors. These observations also suggest that the relevant semantic information identified by \FCAN in the user and item review documents facilitates an accurate interpretation about the recommendation decisions.

\begin{table*}
\scriptsize
\centering
\caption{Highlighted words by attentive weights in the review documents of two user-item pairs for a sampled user $u$. }
\begin{tabular}{l||l}
\hline\hline
Pair$_1(user_u, item_1, 5.0)$ & Pair$_2(user_u, item_2, 3.0)$\\
\hline
$user_u$'s review document: & $user_u$'s review document: \\
\hline
\tabincell{m{0.9\columnwidth}}{So good that I bought another one. \hl{Love} the \hl{heavy cord} and \hl{gold} connectors. Bass sounds great. I just learned last night how to coil them up. I guess I should \hl{read instructions} more \hl{carefully}. But no \hl{harm} done, still works great! \\ \\
$\cdots$, then it worked like a \hl{charm}! \hl{Glad} I \hl{got} them, much less \hl{hassle tuning up} for a \hl{gig}.\\\\
This \hl{tiny} \hlgol{little amp} is \hlgol{great} for \hlgol{practicing outside} or in a canoe or while hang gliding. But it is not very good at parties. It cannot compete with the noise level that 20 people in conversation can make. It is \hl{cheap}, so you should just get one. I am having great \hlgol{fun} experimenting around with different settings, \hl{instruments}, and \hl{power feed voltage} levels.  (see the danelectro power supply for this thing, really \hl{cool}!) \\ \\
$\cdots$ it would be \hlgol{nice} if it remembered what channel it was last on so when you power cycle it doesn't switch on you. I mostly use it to feed one amp from two different instruments. $\cdots$ \\\\
 I use this cable to patch a preamp to an amp.  it works really well.  having the two different \hl{style} of ends gives you \hlgol{good options} when \hlgol{putting} a \hlgol{system together}. \\\\
 \hlgol{Nice} \hl{cable}, could be a little heavier. I managed to pull the wires apart on the straight plug, $\cdots$ \\\\ $\cdots$  it is easy and \hlgol{fun} to use! I stopped messign with the reverb in my amp and now only use this for a much nice sounds from my guitar. I \hl{actually bought} it to try some experiments with my harmonica, but those were dismal failures.  Big \hl{win} for the guitar though!$\cdots$\\\\ \centering{$\cdots\cdots$}}  & \tabincell{m{0.9\columnwidth}}{So good that I bought another one. \hl{Love} the \hl{heavy cord} and \hl{gold connectors}. Bass sounds great. I just learned last night how to coil them up. I guess I should \hl{read instructions} more \hl{carefully}. But no \hlgol{harm} done, still works great! \\\\ $\cdots$, then it worked like a charm! \hl{Glad} I got them, much less \hlgol{hassle tuning up} for a \hlgol{gig}. \\\\ This tiny little amp is great for practicing outside or in a canoe or while hang gliding. But it is not very good at parties. It cannot compete with the \hl{noise level} that \hl{20 people in conversation} can make. It is cheap, so you should just get one. I am having great fun experimenting around with different settings, \hl{instruments}, and \hlgol{power} \hl{feed voltage} levels. (see the danelectro power supply for this thing, really cool!)\\\\
 $\cdots$ it would be \hl{nice} if it \hl{remembered} what \hl{channel} it was last on so when you power cycle it doesn't switch on you.  I mostly use it to feed one \hlgol{amp} from two different \hl{instruments}.$\cdots$ \\\\ I use this cable to patch a preamp to an amp.  it works really well. having the two different \hl{style} of ends gives you \hl{good options} when \hl{putting} a \hl{system together}. \\\\
 \hlgol{Nice} \hl{cable}, could be a little heavier.  I managed to pull the \hl{wires} apart on the \hl{straight plug}, $\cdots$ \\\\
 $\cdots$ it is easy and fun to use! I stopped messign with the reverb in my amp and now only use this for a much nice sounds from my guitar. I \hl{actually} \hlgol{bought} it to \hl{try} some experiments with my harmonica, but those were dismal failures. Big win for the guitar though! \\\\ \centering{$\cdots\cdots$}} \\
\hline\hline
$item_1$'s review document: & $item_2$'s review document: \\
\hline
\tabincell{m{0.9\columnwidth}}{$\cdots$ These are really great \hl{bang} for the \hlgol{buck}, however, like many people said, they do not work well for all my \hl{pedals}, round ends don't \hlgol{fit} in everything.I mean these are cables. \\\\ \hl{Defiantly} a \hl{space saver}. I can put all of my effects side to side ad hook em up with these patch cables. It's very \hlgol{convenient} that they are \hlgol{angled} like so, for \hlgol{low profile} \hl{fits}. My only complaint is that they easily fall out of the socket if you jostle them too much. \\\\
These are very good \hl{quality} in spite of their \hl{reasonable price}. I \hl{love the flat}, \hlgol{low profile}, \hl{right angle} plugs. My only complaint is I needed some in varying lengths $\cdots$ \\\\
I'm running 6 pedals and \hl{no issues like crackling} or \hlgol{cutting out}.One thing that is important to note is that I'm not \hl{hard} on my \hl{gear}. I'm a home noodler, so I can't say if these patches can take road/stage abuse and lots of setup/tear-down cycles. They seem pretty \hl{buff} and \hl{capable}, but I can only report what I've experienced.\\\\
$\cdots$ the design means that you can \hl{save space} in arranging your effects \hl{pedals}, with \hl{enough length} and \hl{flexibility} to arrange them in a \hlgol{semicircle} if, like me, you have a lot of pedals.\\\\ I use these patch cables on my pedal board and they \hlgol{save valuable space}. They work well and I haven't had any problems with any of them shorting out or making static or other noises. I \hl{highly recommend} them. Plus, they \hl{look nice}, unlike the various colors that come with the bulkier, cheaper 1' cords. \\\\ $\cdots$ Simple. Solid. \hl{Compact}. \hlgol{Convenient}. \hl{Audio quality} is \hlgol{unchanged} from my no name metal patch cables. $\cdots$ \\\\ \centering{$\cdots\cdots$}} & \tabincell{m{0.9\columnwidth}}{As with all \hl{Boss}/\hl{Roland} products, the \hl{Boss} FS-6 performs as designed. I purchased it to \hl{control} both my \hl{Boss ME 20 Multi} effects pedal and my Fender Blues Jr amp $\cdots$ Everything \hl{works exactly as planned}. The FS-6 rugged, compact (fits right into my into ME 20 carrying case) and easy to use.I \hl{read} a few \hl{reviews} where people were \hl{complaining} about \hl{batteries} $\cdots$ something to \hlgol{complain} about. As with most \hl{guitar players} I don't use \hl{batteries} in my \hl{effects pedals} I use an 9 volt AC/DC adapter. Visual Sound offers the ``One Spot" AC/DC adapter kit that has a 9 \hl{volt battery} snap connector \hl{adapter}.  I removed the \hl{battery} cover, $\cdots$ I then put rubber feet (purchased at ACE Hardware) on the bottom to \hl{allow} the \hl{snap connector} cable $\cdots$ \hl{battery} problem solved $\cdots$ The possibilities are \hl{limited} only to your \hl{imagination}. It is \hlgol{somewhat expensive} but it is a Great \hl{Product}!! If you need a duel switch then I highly recommend purchasing the FS-6. \\\\ I needed a footswitch for a variety of applications to include an acoustic amp, guitar amps, and other guitar/audio gear. Built like a \hl{tank} and \hl{completely configurable}. Well done again, Roland! \\\\
$\cdots$ I use the to go up and down through my \hl{banks} I have \hl{programmed} on the ME-70. \hl{Build like} a \hl{tank}. Simply to use. Simply to hook up. I hear it can be used on other \hl{Boss} and \hl{Roland} gear.\\\\ being new to guitar this is just way too much fun, makes my playing acutally sound like I may know what I am doing, \hl{easy} to \hl{setup} and easy to use. \\\\ \centering{$\cdots\cdots$} }\\
\hline\hline
$user_u$'s review for $item_1$: & $user_u$'s review for $item_2$:\\
\hline
\tabincell{m{0.9\columnwidth}}{\hlred{compact} heads allow more pedals to be loaded onto your overpriced pedal board.  the \hlred{quality is good}. I \hlred{like} how they are \hlred{easy to take apart and modify and put together} again.  very \hlred{handy} for making it all just right.  I will buy these again for sure.} & \tabincell{m{0.9\columnwidth}}{\hlred{Neither delighted nor disappointed}.  I have \hlred{not found it to be intuitive to use} and \hlred{the battery compartment} is a joke, but otherwise the \hlred{construction is good}.  I dont use it very much because I havent really figured out what I want to do with it, and it is \hlred{too big to fit} in my briefcase full of blues harps, so I have the behringer A/B switch doing the job there.}\\
\hline
\end{tabular} \label{tbl:case}
\end{table*}

\section{Conclusion}\label{sec:con}
In this paper, we propose a novel fused context-aware neural model to learn user-item representations for rating prediction, named \FCAN. Both reviews and user-item rating scores are well exploited in \FCAN. By learning a representation for a user-item pair based on their individual characteristics and their interactions together, \FCAN yields a better understanding of user rating behaviors. The experimental results show that \FCAN consistently outperforms the existing state-of-the-art alternatives over five real-world datasets. Besides, the attention mechanism utilized by \FCAN for review processing enables us to further provide semantic interpretations about recommendation decisions. Inspired by the promising performance delivered by RBLT and the case studies conducted for \FCAN, we plan to incorporate sentiment factors into \FCAN to enhance the rating prediction. Also, the two learning components in \FCAN are just coupled late in a linear fusion manner. As a part of future work, we plan to incorporate the two information sources (\ie reviews and rating scores) into a unified neural model that jointly derives a latent representation for a user-item pair.

\section*{Acknowledgment}\label{sec:ack}
This research was supported by National Natural Science Foundation of China (No.61502344, No.61472287, No.61772377), Natural Scientific Research Program of Hubei Province (No.2017CFB502, No.2017CFA007), Natural Scientific Research Program of Wuhan University (No.2042017kf0225, No.2042016kf0190), Academic Team Building Plan for Young Scholars from Wuhan University (No.Whu2016012). We gratefully acknowledge the support of NVIDIA Corporation with the donation of the Titan X GPU used for this research. Chenliang Li is the corresponding author.

\bibliographystyle{IEEEtran}
\bibliography{ref}

\begin{thebibliography}{10}
\providecommand{\url}[1]{#1}
\csname url@samestyle\endcsname
\providecommand{\newblock}{\relax}
\providecommand{\bibinfo}[2]{#2}
\providecommand{\BIBentrySTDinterwordspacing}{\spaceskip=0pt\relax}
\providecommand{\BIBentryALTinterwordstretchfactor}{4}
\providecommand{\BIBentryALTinterwordspacing}{\spaceskip=\fontdimen2\font plus
\BIBentryALTinterwordstretchfactor\fontdimen3\font minus
  \fontdimen4\font\relax}
\providecommand{\BIBforeignlanguage}[2]{{%
\expandafter\ifx\csname l@#1\endcsname\relax
\typeout{** WARNING: IEEEtran.bst: No hyphenation pattern has been}%
\typeout{** loaded for the language `#1'. Using the pattern for}%
\typeout{** the default language instead.}%
\else
\language=\csname l@#1\endcsname
\fi
#2}}
\providecommand{\BIBdecl}{\relax}
\BIBdecl

\bibitem{kdd11:ctr}
C.~Wang and D.~M. Blei, ``Collaborative topic modeling for recommending
  scientific articles,'' in \emph{Proceedings of the 17th ACM SIGKDD
  International Conference on Knowledge Discovery and Data Mining}, 2011, pp.
  448--456.

\bibitem{recsys13:hfht}
J.~J. McAuley and J.~Leskovec, ``Hidden factors and hidden topics:
  understanding rating dimensions with review text,'' in \emph{Proceedings of
  the 7th ACM Conference on Recommender Systems}, 2013, pp. 165--172.

\bibitem{aaai14:bao}
Y.~Bao, H.~Fang, and J.~Zhang, ``Topicmf: Simultaneously exploiting ratings and
  reviews for recommendation,'' in \emph{Proceedings of the Twenty-Eighth AAAI
  Conference on Artificial Intelligence}, 2014, pp. 2--8.

\bibitem{ijcai16:rblt}
Y.~Tan, M.~Zhang, Y.~Liu, and S.~Ma, ``Rating-boosted latent topics:
  Understanding users and items with ratings and reviews,'' in
  \emph{Proceedings of the Twenty-Fifth International Joint Conference on
  Artificial Intelligence}, 2016, pp. 2640--2646.

\bibitem{jmlr03:lda}
D.~M. Blei, A.~Y. Ng, and M.~I. Jordan, ``Latent dirichlet allocation,''
  \emph{Journal of Machine Learning Research}, vol.~3, pp. 993--1022, 2003.

\bibitem{nips00:nmf}
D.~D. Lee and H.~S. Seung, ``Algorithms for non-negative matrix
  factorization,'' in \emph{Proceedings of the Advances in Neural Information
  Processing Systems 13}, 2000, pp. 556--562.

\bibitem{recsys16:convMf}
D.~H. Kim, C.~Park, J.~Oh, S.~Lee, and H.~Yu, ``Convolutional matrix
  factorization for document context-aware recommendation,'' in
  \emph{Proceedings of the 10th ACM Conference on Recommender Systems}, 2016,
  pp. 233--240.

\bibitem{wsdm17:deepconn}
L.~Zheng, V.~Noroozi, and P.~S. Yu, ``Joint deep modeling of users and items
  using reviews for recommendation,'' in \emph{Proceedings of the Tenth ACM
  International Conference on Web Search and Data Mining}, 2017, pp. 425--434.

\bibitem{recsys17:transnet}
R.~Catherine and W.~W. Cohen, ``Transnets: Learning to transform for
  recommendation,'' in \emph{Proceedings of the 11th ACM Conference on
  Recommender Systems}, 2017, pp. 288--296.

\bibitem{recsys17:dattn}
S.~Seo, J.~Huang, H.~Yang, and Y.~Liu, ``Interpretable convolutional neural
  networks with dual local and global attention for review rating prediction,''
  in \emph{Proceedings of the Eleventh ACM Conference on Recommender Systems},
  2017, pp. 297--305.

\bibitem{icdm10:fm}
S.~Rendle, ``Factorization machines,'' in \emph{Proceedings of the 10th IEEE
  International Conference on Data Mining}, 2010, pp. 995--1000.

\bibitem{advai09:survyCF}
X.~Su and T.~M. Khoshgoftaar, ``A survey of collaborative filtering
  techniques,'' \emph{Advances in Artificial Intelligence}, vol. 2009, pp.
  421\,425:1--421\,425:19, 2009.

\bibitem{nips07:pmf}
R.~Salakhutdinov and A.~Mnih, ``Probabilistic matrix factorization,'' in
  \emph{Proceedings of the Advances in Neural Information Processing Systems
  20}, 2007, pp. 1257--1264.

\bibitem{computer09:mf}
Y.~Koren, R.~M. Bell, and C.~Volinsky, ``Matrix factorization techniques for
  recommender systems,'' \emph{{IEEE} Computer}, vol.~42, no.~8, pp. 30--37,
  2009.

\bibitem{kdd08:neighbor1}
Y.~Koren, ``Factorization meets the neighborhood: a multifaceted collaborative
  filtering model,'' in \emph{Proceedings of the 14th ACM SIGKDD International
  Conference on Knowledge Discovery and Data Mining}, 2008, pp. 426--434.

\bibitem{www17:ncf}
X.~He, L.~Liao, H.~Zhang, L.~Nie, X.~Hu, and T.~Chua, ``Neural collaborative
  filtering,'' in \emph{Proceedings of the 26th International Conference on
  World Wide Web}, 2017, pp. 173--182.

\bibitem{icml07:rbmcf}
R.~Salakhutdinov, A.~Mnih, and G.~E. Hinton, ``Restricted boltzmann machines
  for collaborative filtering,'' in \emph{Proceedings of the 24th International
  Conference on Machine Learning}, 2007, pp. 791--798.

\bibitem{icml13:cfrbm}
K.~Georgiev and P.~Nakov, ``A non-iid framework for collaborative filtering
  with restricted boltzmann machines,'' in \emph{Proceedings of the 30th
  International Conference on Machine Learning}, 2013, pp. 1148--1156.

\bibitem{is04:itembasedrec}
M.~Deshpande and G.~Karypis, ``Item-based top-\emph{N} recommendation
  algorithms,'' \emph{ACM Transactions on Information Systems}, vol.~22, no.~1,
  pp. 143--177, 2004.

\bibitem{recsys14:rmr}
G.~Ling, M.~R. Lyu, and I.~King, ``Ratings meet reviews, a combined approach to
  recommend,'' in \emph{Proceedings of the 8th ACM Conference on Recommender
  Systems}, 2014, pp. 105--112.

\bibitem{kdd15:cdl}
H.~Wang, N.~Wang, and D.~Yeung, ``Collaborative deep learning for recommender
  systems,'' in \emph{Proceedings of the 21th ACM SIGKDD International
  Conference on Knowledge Discovery and Data Mining}, 2015, pp. 1235--1244.

\bibitem{jmlr10:sdae}
P.~Vincent, H.~Larochelle, I.~Lajoie, Y.~Bengio, and P.~Manzagol, ``Stacked
  denoising autoencoders: Learning useful representations in a deep network
  with a local denoising criterion,'' \emph{Journal of Machine Learning
  Research}, vol.~11, pp. 3371--3408, 2010.

\bibitem{ijcai16:cmle}
W.~Zhang, Q.~Yuan, J.~Han, and J.~Wang, ``Collaborative multi-level embedding
  learning from reviews for rating prediction,'' in \emph{Proceedings of the
  Twenty-Fifth International Joint Conference on Artificial Intelligence},
  2016, pp. 2986--2992.

\bibitem{corr14:attention}
D.~Bahdanau, K.~Cho, and Y.~Bengio, ``Neural machine translation by jointly
  learning to align and translate,'' \emph{CoRR}, vol. abs/1409.0473, 2014.

\bibitem{sigir17:acf}
J.~Chen, H.~Zhang, X.~He, L.~Nie, W.~Liu, and T.~Chua, ``Attentive
  collaborative filtering: Multimedia recommendation with item- and
  component-level attention,'' in \emph{Proceedings of the 40th International
  ACM SIGIR Conference on Research and Development in Information Retrieval},
  2017, pp. 335--344.

\bibitem{icdm08:impltrec}
R.~Pan, Y.~Zhou, B.~Cao, N.~N. Liu, R.~M. Lukose, M.~Scholz, and Q.~Yang,
  ``One-class collaborative filtering,'' in \emph{Proceedings of the 8th IEEE
  International Conference on Data Mining}, 2008, pp. 502--511.

\bibitem{sigir16:fmif}
X.~He, H.~Zhang, M.~Kan, and T.~Chua, ``Fast matrix factorization for online
  recommendation with implicit feedback,'' in \emph{Proceedings of the 39th
  International ACM SIGIR Conference on Research and Development in Information
  Retrieval}, 2016, pp. 549--558.

\bibitem{kdd17:dadm}
X.~Wang, L.~Yu, K.~Ren, G.~Tao, W.~Zhang, Y.~Yu, and J.~Wang, ``Dynamic
  attention deep model for article recommendation by learning human editors'
  demonstration,'' in \emph{Proceedings of the 23rd ACM SIGKDD International
  Conference on Knowledge Discovery and Data Mining}, 2017, pp. 2051--2059.

\bibitem{tacl16:yin}
W.~Yin, H.~Sch{\"{u}}tze, B.~Xiang, and B.~Zhou, ``{ABCNN:} attention-based
  convolutional neural network for modeling sentence pairs,''
  \emph{Transactions of the Association for Computational Linguistics}, vol.~4,
  pp. 259--272, 2016.

\bibitem{acl16:wang}
B.~Wang, K.~Liu, and J.~Zhao, ``Inner attention based recurrent neural networks
  for answer selection,'' in \emph{Proceedings of the 54th Annual Meeting of
  the Association for Computational Linguistics}, 2016.

\bibitem{aaai16:wan}
S.~Wan, Y.~Lan, J.~Guo, J.~Xu, L.~Pang, and X.~Cheng, ``A deep architecture for
  semantic matching with multiple positional sentence representations,'' in
  \emph{Proceedings of the Thirtieth AAAI Conference on Artificial
  Intelligence}, 2016, pp. 2835--2841.

\bibitem{corr16:attention}
C.~N. dos Santos, M.~Tan, B.~Xiang, and B.~Zhou, ``Attentive pooling
  networks,'' \emph{CoRR}, vol. abs/1602.03609, 2016.

\bibitem{acl17:cane}
C.~Tu, H.~Liu, Z.~Liu, and M.~Sun, ``{CANE:} context-aware network embedding
  for relation modeling,'' in \emph{Proceedings of the 55th Annual Meeting of
  the Association for Computational Linguistics}, 2017, pp. 1722--1731.

\bibitem{lecture12:recommend}
A.~Smola, ``Recommender systems lecture,''
  \url{http://alex.smola.org/teaching/berkeley2012/slides/8_Recommender.pdf},
  2012.

\bibitem{jmlr11:nlp}
R.~Collobert, J.~Weston, L.~Bottou, M.~Karlen, K.~Kavukcuoglu, and P.~P. Kuksa,
  ``Natural language processing (almost) from scratch,'' \emph{Journal of
  Machine Learning Research}, vol.~12, pp. 2493--2537, 2011.

\bibitem{acl14:cnn1}
N.~Kalchbrenner, E.~Grefenstette, and P.~Blunsom, ``A convolutional neural
  network for modelling sentences,'' in \emph{Proceedings of the 52nd Annual
  Meeting of the Association for Computational Linguistics,}, 2014, pp.
  655--665.

\bibitem{emnlp14:cnn2}
Y.~Kim, ``Convolutional neural networks for sentence classification,'' in
  \emph{Proceedings of the 2014 Conference on Empirical Methods in Natural
  Language Processing}, 2014, pp. 1746--1751.

\bibitem{nips:lihangcnn}
B.~Hu, Z.~Lu, H.~Li, and Q.~Chen, ``Convolutional neural network architectures
  for matching natural language sentences,'' in \emph{Proceedings of the
  Advances in Neural Information Processing Systems 27}, 2014, pp. 2042--2050.

\bibitem{lecture12rmsp}
T.~Tieleman and G.~Hinton, ``Lecture 6.5-rmsprop: Divide the gradient by a
  running average of its recent magnitude,'' \emph{COURSERA: Neural networks
  for machine learning}, vol.~4, no.~2, pp. 26--31, 2012.

\bibitem{jmlr14:dropout}
N.~Srivastava, G.~E. Hinton, A.~Krizhevsky, I.~Sutskever, and R.~Salakhutdinov,
  ``Dropout: a simple way to prevent neural networks from overfitting,''
  \emph{Journal of Machine Learning Research}, vol.~15, no.~1, pp. 1929--1958,
  2014.

\bibitem{www16:datasets}
R.~He and J.~McAuley, ``Ups and downs: Modeling the visual evolution of fashion
  trends with one-class collaborative filtering,'' in \emph{Proceedings of the
  25th International Conference on World Wide Web}, 2016, pp. 507--517.

\end{thebibliography}

\begin{IEEEbiography}{Libing Wu} received the B.Sc. and M.Sc. degrees
from Central China Normal University, Wuhan,
China, in 1994 and 2001, respectively, and the Ph.D.
degree from Wuhan University, Wuhan, in 2006, all
in computer science. He was a Visiting Scholar with
the Advanced Networking Laboratory, University of
Kentucky, USA, in 2011. He is currently a Professor
with the School of Computer Science, Wuhan
University. His research interests include wireless
sensor networks, network management, and distributed
computing.
\end{IEEEbiography}

\begin{IEEEbiography}{Cong Quan} is currently an Ph.D student at Wuhan University, under the supervision of Dr. Chengliang Li and Dr. Libing Wu. He receive MSc Degree from the Chinese University of Hong Kong in 2014. His research interests lie in information retrieval, recommendation systems and data mining.
\end{IEEEbiography}

\begin{IEEEbiography}{Chenliang Li}
received PhD from Nanyang Technological University, Singapore, in 2013. Currently, he is an Associate Professor at State Key Laboratory of Software Engineering, School of Computer Science, Wuhan University, China. His research interests include information retrieval, text/web mining, data mining and natural language processing. Most of his papers appear in SIGIR, CIKM, TKDE, TOIS and JASIST. He is a co-recipient of Best Student Paper Award Honorable Mention in ACM SIGIR 2016, and serves now as an editorial board member of JASIST. 
\end{IEEEbiography}

\begin{IEEEbiography}{Qian Wang} received the B.S. degree
from Wuhan University, Wuhan, China, in 2003,
the M.S. degree from the Shanghai Institute of
Microsystem and Information Technology, Chinese
Academy of Sciences, Beijing, China, in 2006,
and the Ph.D. degree from the Illinois Institute of
Technology, Chicago, IL, USA, in 2012, all in electrical
engineering. He is currently a Professor with the School of
Computer Science, Wuhan University. His current
research interests include wireless network security
and privacy, cloud computing security, and applied cryptography. He is an expert under “1000 Young Talents Program” of China and an ACM member.
\end{IEEEbiography}

\begin{IEEEbiography}{Bolong Zheng} is currently a Research Professor at the School of Data and Computer Science, Sun Yat-sen University. He received his PhD from University of Queensland in 2017. He received bachelor and master degrees in computer science from Huazhong University of Science and Technology, Wuhan, China, in 2011 and 2013 respectively. His research interests include spatial database, spatial keyword search and graph database.
\end{IEEEbiography}

\end{document}